\documentclass[10pt,journal,compsoc]{IEEEtran}

\ifCLASSOPTIONcompsoc
  \usepackage[nocompress]{cite}
\else
  \usepackage{cite}
\fi
\ifCLASSINFOpdf
\else
\fi

\usepackage{subfigure}
\usepackage{amsmath}
\usepackage{amsfonts}
\usepackage{graphicx}  
\usepackage{amssymb}
\usepackage{dsfont}
\usepackage{multirow}
\usepackage[ruled,vlined]{algorithm2e}
\SetKw{KwBy}{by}
\SetKw{Kwand}{and}

\usepackage[hyphens]{url}
\usepackage[hidelinks]{hyperref}
\hypersetup{breaklinks=true}
\usepackage{cite}

\newcommand{\bmark}[1]{#1}

\begin{document}
%
\title{In Situ Data Summaries for Flexible Feature Analysis in Large-Scale Multiphase Flow Simulations}
%
%
%
%

\author{Soumya~Dutta,~\IEEEmembership{Member,~IEEE}, 
	Terece~L.~Turton,~\IEEEmembership{Member,~IEEE}, 
	David~H.~Rogers, 
	Jordan~M.~Musser, 
	James~P.~Ahrens, 
	and Ann~S.~Almgren
\IEEEcompsocitemizethanks{\IEEEcompsocthanksitem S. Dutta is with Los Alamos National Laboratory.\protect\\
E-mail: sdutta@lanl.gov.
\IEEEcompsocthanksitem S. Dutta, T. Turton, D. Rogers, and J. Ahrens are with Los Alamos National Laboratory.
\IEEEcompsocthanksitem J. Musser is with National Energy Technology Laboratory.
\IEEEcompsocthanksitem A. Almgren is with Lawrence Berkeley National Laboratory.
}
\thanks{Manuscript received November xx.xxxx ; revised xxxx.xxxx}}

%
%

\markboth{Under Review at IEEE TVCG, 2022}%
{Shell \MakeLowercase{\textit{et al.}}: Bare Advanced Demo of IEEEtran.cls for IEEE Computer Society Journals}
%



\IEEEtitleabstractindextext{%
\begin{abstract}
The study of multiphase flow is essential for understanding the complex interactions of various materials. In particular, when designing chemical reactors such as fluidized bed reactors (FBR), a detailed understanding of the hydrodynamics is critical for optimizing reactor performance and stability \cite{BubbleTree,Boyce_single_bubble}. An FBR allows experts to conduct different types of chemical reactions involving multiphase materials, especially interaction between gas and solids. During such complex chemical processes, formation of void regions in the reactor, generally termed as \textit{bubbles}, is an important phenomenon. Study of these bubbles has a deep implication in predicting the reactor's overall efficiency. But physical experiments needed to understand bubble dynamics are costly and non-trivial due to the technical difficulties involved and harsh working conditions of the reactors \cite{BubbleTree}. Therefore, to study such chemical processes and bubble dynamics, a state-of-the-art massively parallel computational fluid dynamics–discrete element model (CFD-DEM), MFIX-Exa \cite{mfix_exa_url,mfix_exa} is being developed for simulating multiphase flows. Despite the proven accuracy of MFIX-Exa in modeling bubbling phenomena, the very-large size of the output data prohibits the use of traditional post hoc analysis capabilities in both storage and I/O time. 
To address these issues and allow the application scientists to explore the bubble dynamics in an efficient and timely manner, we have developed an end-to-end visual analytics pipeline that enables in situ detection of bubbles using statistical techniques, followed by a flexible and interactive visual exploration of bubble dynamics in the post hoc analysis phase. The proposed method enables interactive analysis and tracking of bubbles, along with quantification of several salient bubble characteristics, enabling experts to understand the bubble interactions in detail and further conceive new hypotheses. Positive feedback from the experts has indicated the efficacy of the proposed approach for exploring bubble dynamics in very-large scale multiphase flow simulations.
\end{abstract}

}

\maketitle
\IEEEdisplaynontitleabstractindextext

%
\IEEEpeerreviewmaketitle

\ifCLASSOPTIONcompsoc
\IEEEraisesectionheading{\section{Introduction}\label{sec:introduction}}
\else
\section{Introduction}
\label{sec:introduction}
\fi

\begin{figure}[tb] 
\centering
\includegraphics[width=0.85\linewidth]{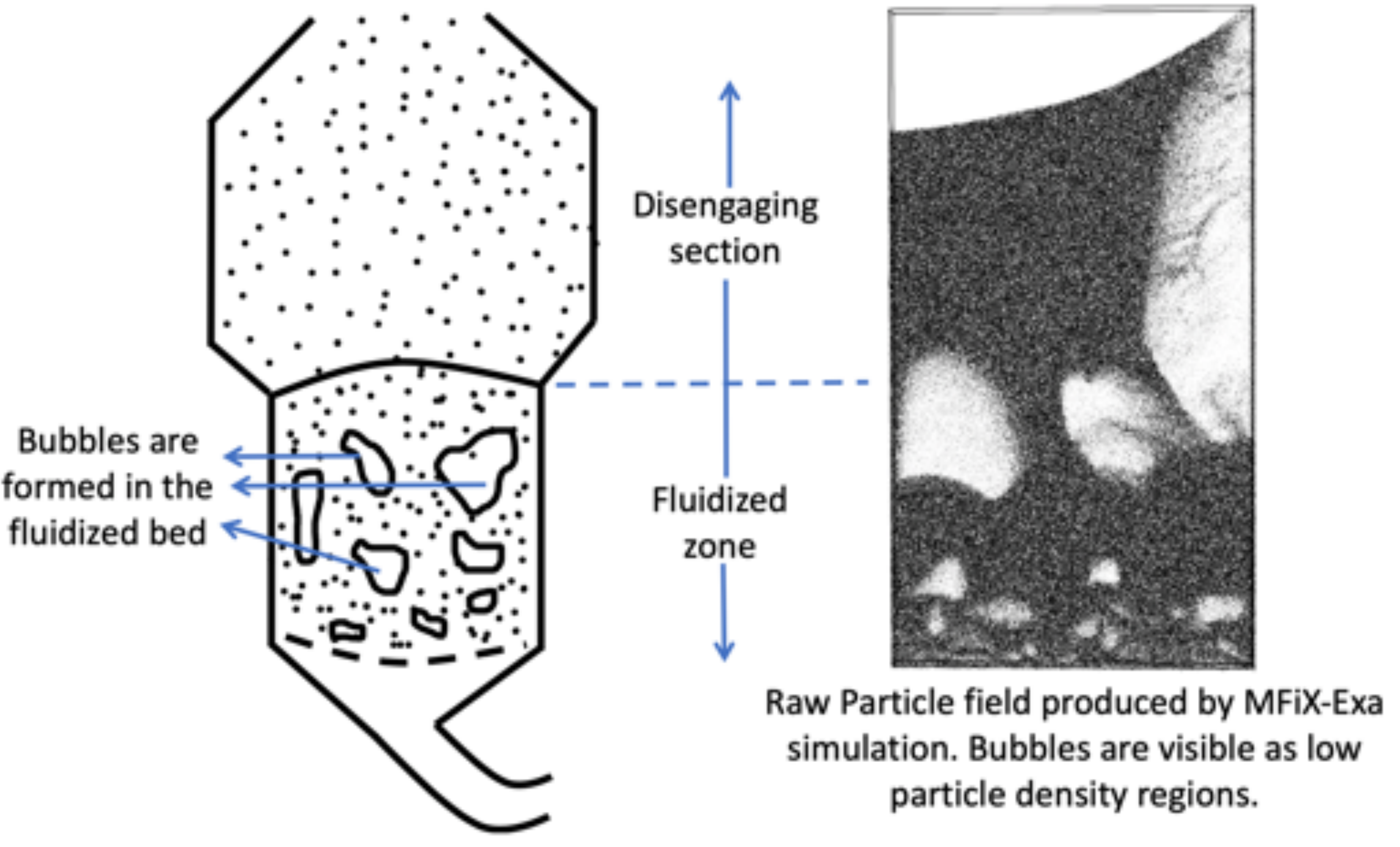}
\caption{A schematic diagram of a fluidized bed where the bubble phenomenon is shown (left). A visualization of a particle field with bubbles produced by MFIX-Exa simulation (right).}
\label{fig:fluidized_bed_structure}
\end{figure}

With recent advancements in  parallel computing capabilities, application scientists are currently building  high-resolution computational models to study the working principles of chemical looping reactors (CLR) by simulating various types of multiphase flows. The study of the temporal evolution and dynamics of bubbles in a fluidized bed is of prime interest. In the fluidization process, bubbles (void regions in the fluidized beds) are formed under certain physical conditions, interacting with each other as shown in Figure~\ref{fig:fluidized_bed_structure}. Understanding the dynamics of such bubbles for these systems is paramount as the formation of large, fast-moving bubbles in fluidized beds causes poor gas/solids mixing, lowering conversion efficiency and stability.

A massively parallel computational fluid dynamics–discrete element model (CFD-DEM) code, MFIX-Exa~\cite{mfix_exa_url} is being developed by the National Energy Technology Laboratory (NETL) to study multiphase flows in detail.  Data generated from MFIX-Exa enables in-depth study of bubble dynamics. However, both computational cost and data size from a single run of the simulation can be large.
As a result, the traditional approach of using post hoc analysis and visualization is becoming prohibitively time-consuming and a limiting factor in the ability to derive insight from the data. The bottleneck stems from I/O speed limitations and extreme output data sizes compared to the ever-increasing computing speed. Storing detailed data for individual time steps (e.g., particle positions and velocities in addition to fluid field quantities) is becoming less viable. 
Scalable and timely visualization and analysis of such data sets poses significant challenges.

The main goal of this work is to develop a practical and reliable solution for domain experts such that they can analyze and visualize bubble dynamics in fluidized beds in a timely manner using large three dimensional data sets generated from MFIX-Exa. The experts are interested in understanding the evolution of bubbles and how various characteristics of bubbles evolve. They wish to explore relationships between bubble characteristics and study their velocity profile as bubbles interact with each other. Given a specific time step, the experts also want to track the bubbles collectively and compare their temporal evolution to generate further insights. Producing these bubble data sets requires lengthy simulations often running for days. The extreme size of the output data is prohibitive to transferring the full data to persistent storage. Therefore, the experts are looking for new solutions that will enable the extraction of important information from the data accurately, leading to a significant amount of storage reduction. Finally, the domain scientists need effective visualization tools that can present the bubble dynamics results via intuitive and simple visual encoding that can be explored, compared, and contrasted interactively and interpreted readily.

In situ analysis techniques provide an attractive solution to address these issues. Since in situ analysis is done as the data is produced in the supercomputer memory, it provides the opportunity to analyze the data in real-time and extract the important information needed to study bubble dynamics. In situ processing can achieve significant data triage and reduction as has been shown recently by many researchers~\cite{Jim_SC_cinema,Hank_SC_2015,triage_in_situ_app_driven,dutta_vis16}. However, to understand bubble dynamics in detail, expert-in-the-loop interactive visual exploration is necessary where experts can query and filter different bubbles and track them to glean insights about the bubbling phenomenon. This process will be more time-intensive. How much of the proposed analysis pipeline should be done in situ and which analyses should be deferred to the post hoc phase needs to be decided carefully to balance the workload between in situ processing and post hoc analysis.

In this work, we present a visual analysis technique that combines both in situ and post hoc analysis strategies resulting in an effective workflow to study  bubble dynamics in fluidized beds using data generated from MFIX-Exa simulations. This work demonstrates the first in situ workflow for MFIX-Exa simulation targeted to address the unavoidable I/O bottleneck and data size explosion challenges. Since the features of interest, the bubbles, referred to as \textit{void regions}, have low particle density, they lack a precise descriptor. In the absence of a precise definition, 
we employ a statistical distribution-based feature detection technique that has been shown effective for the detection of uncertain features~\cite{dutta_slic_pvis,dutta_vis15,dist_huang,Lundstrom_2006}. To reliably detect bubbles, we first compute the particle density field from unstructured raw particles and then model the density field as a homogeneously partitioned distribution field. Finally, the statistical similarity of each partition to that of the target bubble feature is quantified. This similarity-based classification of the particle density field results in a new scalar field called the bubble similarity field (BSF), where each point indicates the possibility of being part of a bubble. Besides the BSFs, to capture the velocity dynamics profile of the bubbles, we also compute another scalar field from the particle velocity data where each point reflects the particle rise velocity. These two derived scalar fields are stored to disk for post hoc analysis.

During the post hoc analysis, the BSFs and the particle rise velocity based scalar fields (PVFs) are explored in detail to analyze and visualize bubble dynamics. Based on the degree of the similarity values, the BSFs are segmented, and connected component analysis is applied to isolate individual bubbles. For each bubble, several salient characteristics such as bubble aspect ratio, rise velocity, volume, and position are computed. To study the temporal evolution of bubbles, an overlap-based tracking algorithm~\cite{Silver96volumetracking,4906833} is applied. An overlap-based tracking algorithm is suitable for our application since we process the data in situ with sufficient temporal frequency to ensure feature overlap between consecutive time steps. Besides tracking individual bubbles, all the bubbles from a specific time step can also be tracked collectively to provide a comprehensive view of the bubble dynamics. The results of computed bubble dynamics and tracking are presented to the experts through several Cinema-based interactive viewers~\cite{cinema_explorer, cinema_view}, which are found to be effective and intuitive in our application study. \bmark{In addition to the Cinema-based interactive tools, the bubble tracking results are also visualized using a custom interactive visualization tool in the 3D spatial domain so that the users can study bubble evolution directly in the 3D domain}. Positive feedback from the domain scientists demonstrates the efficacy of our proposed technique in analyzing bubble dynamics in multiphase flow simulations. Therefore, our contributions in this work are threefold:

\begin{enumerate}
\item We develop a novel in situ analysis workflow for MFIX-Exa simulations and employ a statistical feature detection algorithm to characterize bubble features in the in situ environment.
\item We propose post hoc analysis and visualization techniques that utilize only the reduced in situ generated feature-aware data summaries to extract, isolate, track, and compare bubbles in an effective and timely manner.
\item We propose a pragmatic, flexible, and scalable end-to-end feature-driven visual analytics workflow for domain experts that enables in-depth exploration of bubble dynamics in multiphase flow simulations. Finally, an in situ performance study is conducted to demonstrate the in situ viability of the proposed technique in handling large-scale particle data sets.
\end{enumerate}

\section{Related Work}
\textbf{In situ analysis and visualization.} The need for in situ data analysis has grown significantly in recent years to address the problems arising from slow disk I/O.
The visualization community has developed several tools for direct in situ rendering of data~\cite{catalyst,visit,ascent,adios,sensei_url,glean}. All these tools can produce high-quality visualization results in situ. However, when an exploratory analysis is needed, conducting the complete analysis in situ would slow down the simulation significantly.
Therefore, a hybrid data analysis paradigm is becoming popular where the data is processed in situ to summarize and extract important information in a compact format and store it to disk for flexible post hoc analysis~\cite{Hank_SC_2015,triage_in_situ_app_driven,in_situ_compression}. This idea has been pursued by many researchers to develop various in situ data summarization algorithms. An image-based in situ data reduction strategy has been shown effective~\cite{Jim_SC_cinema,explorable_images}. Statistical distribution-based in situ data summarization techniques have become popular in recent years due to their compactness and flexibility~\cite{dutta_vis16,dutta_slic_pvis,kochih_spatialdist,hazarika_codda,insitu_histogram}. 
Statistical downsampling for in situ data reduction has also been explored~\cite{tzu_sampling,isav_sampling,Woodring_sampling,dutta2019multivariate,kochih_superres,void_cluster_sampling}. 
An in situ trigger infrastructure has been developed by Larsen et al. \cite{insitu_trigger}. In this current work, we employ an in situ distribution-based feature detection strategy for efficiently classifying bubble features from particle data sets.

\textbf{Statistical feature analysis.} As the complexity of scientific features has grown, the use of statistical techniques for exploring features in scientific data sets has gained popularity among visualization researchers. Integral histograms were used by scientists to explore local features in scientific data sets~\cite{Abon_Pvis14,lee_wavelet}. To advance the query-driven visualization capabilities, Gosink et al.~\cite{gosink_dist_qdv} used distribution functions for feature analysis. For an enhanced understanding of features in data sets, Johnson and Huang~\cite{dist_huang} allowed querying on distributions via fuzzy feature matching. Features in ensemble data sets were explored using generalized boxplot-based visualizations~\cite{boxplot,contour_boxplot}. Thompson et al.~\cite{Thompson2011} introduced the idea of hixel, which enabled data summarization as well as the preservation of features in the reduced data sets. Local distribution-based feature extraction and searching was also explored by several researchers~\cite{CGF12620,dutta_slic_pvis,kochih_spatialdist,Lundstrom_2006,GMM_timevarying_tvcg,Liu2012}. In our work, we follow a similar approach where the data is modeled using local region-wise distribution models, and then bubble features are detected based on statistical similarity.

\textbf{Feature tracking.} Feature tracking is considered one of the fundamental visualization tasks for analyzing time-varying data features. A significant amount of research has been done on developing different feature tracking algorithms. Silver et al.~\cite{Silver96volumetracking,Silver1998} proposed one of the earliest feature tracking algorithms that used overlapp for feature correspondence. An attribute-based feature tracking was proposed by Samatanay et al.~\cite{Samtaney1994}. Earth mover's distance was used for feature tracking by Ji and Shen~\cite{Ji_featuretracking}. Higher-dimensional isosurface based feature tracking was also proposed by Ji et al.~\cite{1250374}. 
For tracking features collectively as a group, Ozer et al. proposed techniques for tracking group dynamics~\cite{Ozer2014,6378982}. Muelder and Ma introduced a predictor-corrector based approach for accurate feature tracking. By utilizing global knowledge from all time steps, a merge tree guided feature tracking was proposed by Saikia and Weinkauf~\cite{global_tracking}. Feature tracking in joint particle/volume data sets was recently proposed by Sauer et al.~\cite{Sauer2014}. Dutta and Shen developed a fuzzy feature tracking algorithm for uncertain features~\cite{dutta_vis15}.  A texture-based feature tracking algorithm was developed by Caban et al.~\cite{Caban}. Topological techniques were also used for characterizing and tracking features in large scientific data sets~\cite{jonas,timo,Widanagamaachchi_tracking}. Tracking of magnetic flux vortices in superconductor data was proposed by Guo et al.~\cite{hanqi_superocnductor}. Feature tracking algorithms were also proposed for tracking features in vector field~\cite{Theisel2003,1372214}. A comprehensive survey of feature tracking can be found in~\cite{PostVHLD03}. In this work, we have used an overlapping-based feature tracking algorithm. By sampling the data frequently during in situ processing, we ensure that an overlapping-based tracking can be applied in our application to solve the feature correspondence.

\section{Application Background and Motivation}
\subsection{Application Background}
Understanding bubble dynamics in fluidized beds is important to scientists studying multiphase flows in order to design efficient, cost-effective chemical looping reactors (CLR). In a typical CLR, oxygen from a solid oxygen carrier, such as metal oxide, is used to combust fossil fuels~\cite{mfix_ecp_url}. In a standard configuration, the solid oxygen carrier circulates between two fluidized beds, preventing the fuel from directly contacting the air~\cite{ABAD2011689}. Large numbers of bubbles can form, causing poor gas/solids mixing and lowering conversion efficiency. Using MFIX-Exa, scientists can analyze and visualize various bubble phenomena occurring in fluidized beds under different physical conditions. MFIX-Exa is one of the simulation codes in the U.S. Department of Energy's Exascale Computing Project (ECP)~\cite{mfix_ecp_url}. It is expected to achieve exascale performance in the near future, enabling high-fidelity simulations. Since storing all the high-resolution raw data is not practical due to I/O limitations, developing an in situ analysis capability is crucial for the domain scientists. MFIX-Exa simulation code is being built using kernels from the existing MFiX project~\cite{mfix_url}, with a software structure redesign using AMReX~\cite{amrex_url,amrex_paper} as its foundation.  AMReX is a block-structured AMR-based software framework designed for building massively parallel applications. MFIX-Exa can produce both mesh and particle data based on the requirements of the application.  This work utilized the unstructured raw particle data to study bubble dynamics phenomena.

\begin{figure}[tb] 
\centering
\includegraphics[width=\linewidth]{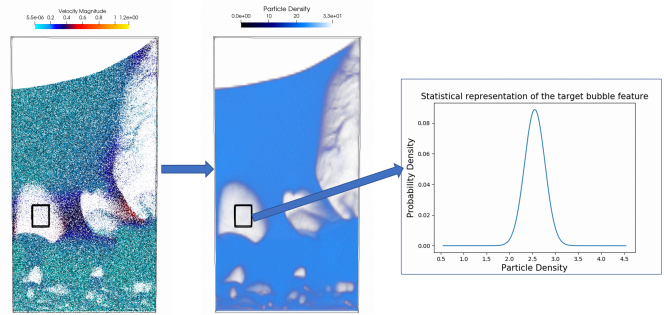}
\caption{Selection of target bubble feature directly from the data.}
\label{fig:feature_select}
\end{figure}
\subsection{Motivation}
Our motivation comes from surveying existing domain capabilities and from discussions with the domain experts about limitations of existing tools for analyzing bubble dynamics. Presently, the experts completely rely on a post-processing workflow with full resolution raw data using tools such as ParaView~\cite{Paraview} or VisIt~\cite{visit}. At current simulation scales, the experts skip on the order of hundreds of time steps when storing data to disk in order to keep the data size tractable, possibly missing meaningful events. 
Given the complex bubble interactions, it is critical to have access to a sufficiently high temporal resolution of the data so that reliable feature tracking can be done with minimal feature correspondence errors. As the scale of the simulation is expected to grow significantly, the frequency of raw data storage will likely go down even more to address I/O limitations. Post-processing workflows will not scale, making in situ processing critical.  Lastly, it can be overwhelming for the domain scientists to analyze and visualize such a large number of bubbles using generic visualization tools as they do not address the specific needs the experts have~\cite{Boyce_single_bubble}.

A recent post-processing pipeline for studying bubbles was conducted by Buchheit et al.~\cite{BubbleTree}.  It used mesh-based data generated from MFiX simulations, the predecessor of MFIX-Exa. Bubbles were identified by a constant threshold in volume fraction field and their contours were visualized and analyzed. Although this technique shows promise, using a fixed threshold to segment bubbles could lead to missing smaller bubbles. Furthermore, while tracking, the intersection between corresponding bubbles was tested by intersecting the bounding boxes which also could result in incorrect correspondence for convex bubbles~\cite{BubbleTree}.  We improve upon these potential shortcomings and utilize the particle data from MFIX-Exa, sampling it frequently in situ to ensure that overlap-based reliable bubble tracking can be applied. The overlap test is done on the 3D bubble features and hence a true overlap is detected. We employ a statistical feature detection technique that generates a bubble similarity field which can be visualized with varying similarity values to illustrate uncertainties associated in bubble detection. We also provide customized visualization tools that effectively show the results of bubble dynamics interactively, enabling the user to 
query specific bubble properties and compare and contrast tracking results.
\begin{figure}[tb] 
\centering
\includegraphics[width=\linewidth]{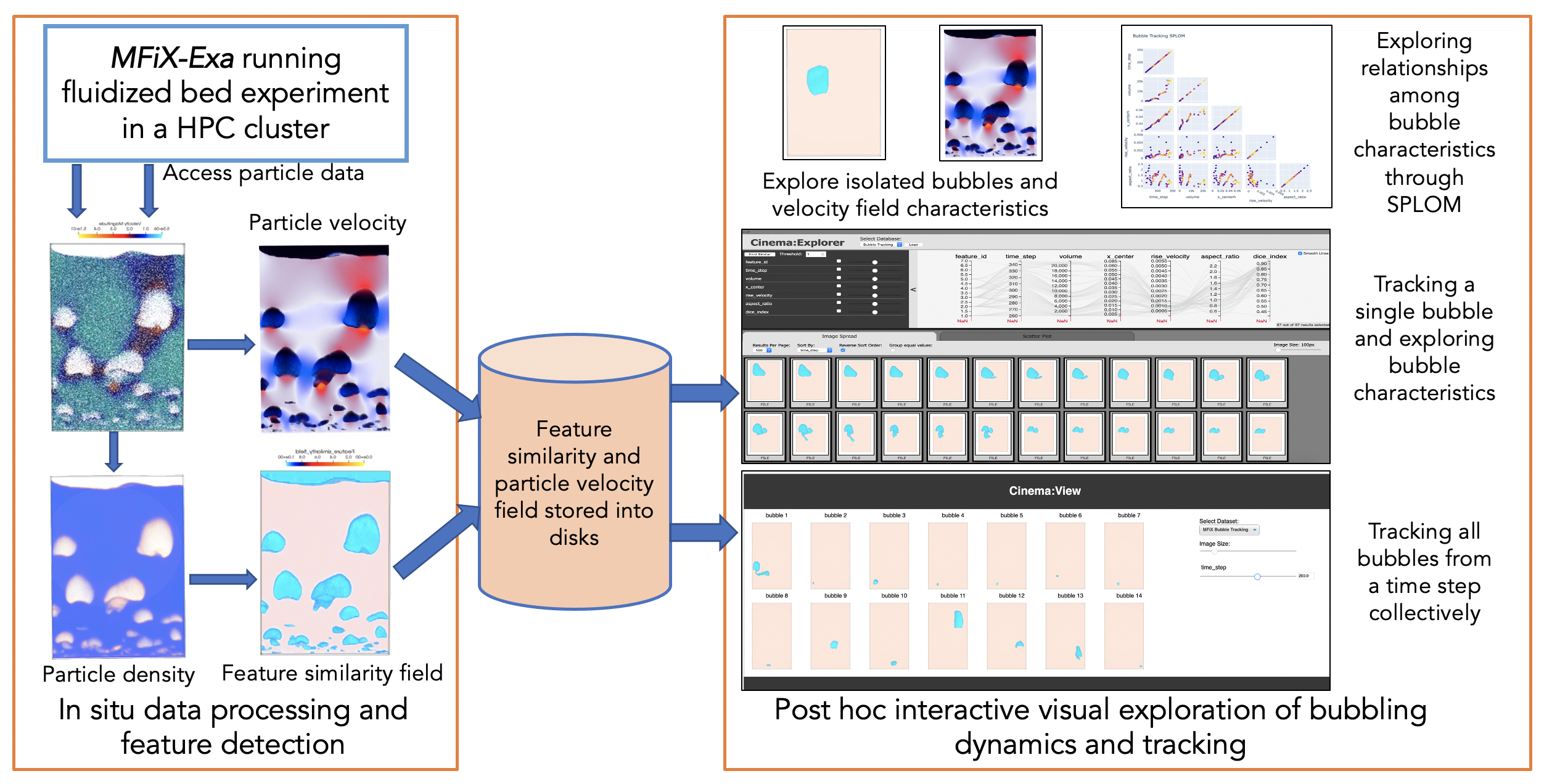}
\caption{An overview of the proposed end-to-end analysis pipeline.}
\label{fig:overview}
\end{figure}

\section{Domain Requirements and Overview}

\subsection{Domain Specific Requirements}
After discussing with the developers of the MFIX-Exa simulation, we have identified several important domain-specific requirements and are listed below:
\begin{enumerate}
\item The storage of full-resolution raw data will not be possible with sufficient temporal fidelity, requiring an in situ analysis driven workflow to extract and summarize the important bubble specific information in situ and store reduced data summaries to disk for flexible post hoc analysis.  
\item The experts need the ability to analyze, visualize, and filter bubbles based on  various bubble characteristics to explore relationships among different bubble characteristics such as bubble volume, shape, rise velocity, etc.
\item The experts want to track evolution of individual bubbles as well as all bubbles from a selected time step collectively to compare and contrast interactions and dynamics. They want to visualize the velocity profile of the particles around  detected bubbles to answer questions such as: Is the rise velocity of bubbles consistent or varying? What is the relationship between bubble volume and rise velocity? How do bubbles merge (or split)?
\item Domain experts also need visualization tools to interactively and intuitively visualize the analysis results and study bubble dynamics in 3D spatial domain.  
 \end{enumerate}

\subsection{Feature Selection and Overview of Our Workflow} \label{feature_selection}
Since the features of interest lack a precise descriptor, we start by having the users select a region of interest directly in the data offline, based on domain expertise,
to highlight a bubble region with low particle density. An example selection is shown in Figure~\ref{fig:feature_select}. The bubble feature is highlighted using a 3D box filter. Next, a particle density field is estimated from the unstructured particle data, shown in the center image of Figure~\ref{fig:feature_select}. 
From this continuous density scalar field, all the grid points that fall within the box filter are collected and a Gaussian distribution is estimated using their particle density  as shown in the rightmost image of Figure~\ref{fig:feature_select}. 
The use of a single Gaussian distribution to model such bubbles is sufficient as the  bubbles in the density field generally form a homogeneous region with minimal density variations.  This keeps the feature description simple and compact.  This Gaussian distribution is used as a statistical representation of the target feature and statistically similar regions to this distribution
can be considered as part of potential bubbles. The use of distributions as a feature template has been shown effective by researchers in the past when a precise feature descriptor is not available \cite{dutta_vis15,dist_based_track_1,prob_tracking_cvpr2003,Lundstrom_2006,Obermaier2013}. 
Distribution-based feature descriptors are not sensitive to object shape, an advantage in dealing with non-rigid, shape-changing objects such as bubbles~\cite{prob_tracking_cvpr2003}.  

\begin{figure*}[tb] 
\centering
\includegraphics[width=0.9\linewidth]{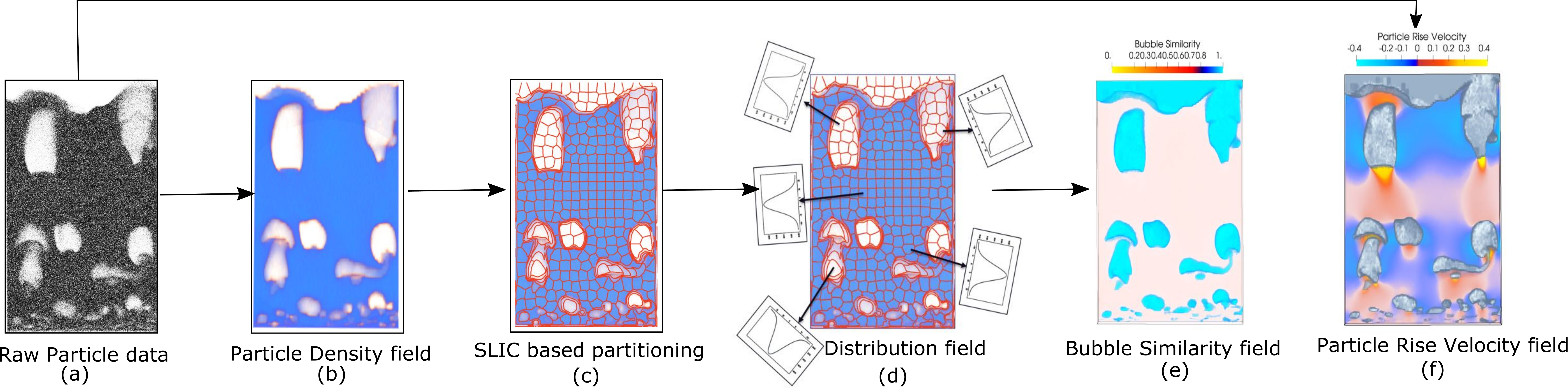}
\caption{Steps of in situ processing for data transformation, modeling, and bubble detection. We start with the particle data (a), and convert it to a particle density field (b). The particle density field is then partitioned into local homogeneous regions (c), and a distribution field is constructed (d). Finally, a bubble similarity field (BSF) (e) is generated by comparing the distribution field with the user-provided target feature. Besides the BSF, a particle rise velocity field (PVF) (f) is also estimated to analyze the particle velocity profiles during post hoc analysis.}
\label{fig:insitu_transformations}
\end{figure*}
With this distribution-based bubble feature template, detection of statistically similar regions in the data is performed in situ. The end-to-end analysis pipeline is shown in Figure~\ref{fig:overview}.  As can be seen, during the in situ processing, the raw particle data is accessed and an intermediate particle density field is computed. Then the particle density field is partitioned into small homogeneous regions using a super-voxel based clustering technique~\cite{slic,slic_ucdavis,dutta_slic_pvis}. Using density values of data points from each partition/super-voxel, a Gaussian distribution is estimated and finally, the statistical similarity of each partition to the template Gaussian distribution is quantified. Then a new bubble similarity field (BSF) is constructed for further bubble analysis. 
An additional scalar field is computed using the particle velocity information to support studies of particle velocity profiles. These two scalar fields are stored to disk for post hoc analysis. In the post hoc analysis, the BSFs are segmented, the bubbles are isolated and their various characteristics are measured. A bubble tracking algorithm is employed to track the evolution of bubbles. All these analysis results are presented to the experts using interactive visualization tools to enable the experts to interact, filter, and track bubbles, exploring how they evolve. By visualizing the particle rise velocity based scalar fields (PVFs), they can understand how the particles around the bubbles behave as the bubbles rise. We also provide summary statistics and scatterplot matrices (SPLOM) to show the relationships between bubble attributes and their time evolution.

\section{In Situ Modeling and Bubble Detection}

In this Section, we describe the in situ data analysis techniques employed to classify bubbles from the raw particle data, generate a bubble similarity field (BSF), and compute a particle rise velocity field (PVF) used during post hoc analysis. While processing data in situ, we transform, model, and finally extract important bubble specific data from the particle data and store such summarized information to disk, significantly reducing I/O. By doing so, we achieve significant data triage and enable flexible and scalable post hoc bubble dynamics analysis. 

\subsection{Particle Field to Density Field Conversion}
Since the raw particle data produced by MFIX-Exa is unstructured and the target bubble features are continuous regions with low particle densities,
we first convert the particle field to a density field. Density estimation is often regarded as a fundamental step necessary for sampling particle fields into a structured continuous representation~\cite{tom_density}. 

We have used a spatial histogram-based technique to group particles into non-overlapping bins and then a density field is finally constructed. As the particles are distributed across various computing nodes
we compute the histogram in the same distributed setting.  A local histogram is first constructed at each processing unit by binning the 3D locations of all particles available to each processor. A 3D histogram is required since we are binning particle locations to estimate spatial particle density. The number of bins and bin-widths on each local processing unit is the same and is estimated from the global bounds. 
Finally, the partial histograms are combined to construct the global density histogram by using a parallel reduction operation over all processing units. Each bin in this global spatial histogram represents particle counts in a local spatial region. The global 3D histogram is transformed into a 3D regular grid based scalar field where each 3D bin center is mapped to a voxel in the regular grid data and the particle count for that bin is assigned as the particle density value at that voxel. Using a spatial histogram-based approach to convert the particle data into a density field can be efficiently estimated in situ, keeping the computational cost low during in situ processing. 
Figure~\ref{fig:insitu_transformations}b shows an example estimated density field for the raw particle field depicted in Figure~\ref{fig:insitu_transformations}a.

\subsection{Homogeneity-guided Density Field Modeling}
Given the lack of a precise descriptor, it is non-trivial to find a consistent hard density threshold value that can be used for isolating low-density regions from the density field. The statistical approach allows classification of the in situ generated density field into a feature similarity field where regions having high feature similarity values can be explored as bubbles. Since the bubbles are contiguous regions with low particle density, to detect them accurately, we follow a local statistical data modeling approach to first convert the density field into a distribution field and then classify the distribution field to produce a bubble similarity field (BSF), based on the user-specified target bubble distribution (see Section~\ref{feature_selection}).

We follow a statistical distribution-based local data modeling technique. Our work advocates an irregular clustering-based data partitioning scheme to maximize homogeneous data values in each partition. The high degree of homogeneity allows for more compact distribution-based data modeling with reduced estimation errors. We have used the Simple Linear Iterative Clustering (SLIC) algorithm~\cite{slic} for producing a homogeneous partitioning of the density field. SLIC is a supervoxel generation algorithm that has been shown to achieve state-of-the-art results both in terms of quality and computational cost. SLIC's superior feature preservation ability in scientific data sets has also been studied~\cite{slic_ucdavis}. 

SLIC is a variant of the local K-means clustering algorithm which works within a predefined local neighborhood while clustering the data. Hence, the total number of distance computations required by SLIC is greatly reduced, resulting in a significant computational speed-up. The algorithm expects the user to specify the approximate size of clusters, $a\times b \times c$. The total number of clusters is estimated and cluster centers are initialized regularly. Since the expected size of a cluster is $a\times b \times c$, the search for similar data points is restricted within a neighborhood $2a\times 2b \times 2c$ around each cluster center~\cite{slic}. This key strategy significantly reduces the total number of distance computations required. The technique iteratively assigns all the data points to the best representative and when the assignments do not change over consecutive iterations, SLIC terminates. In order to keep a balanced contribution from the data similarity and spatial proximity while assigning data points to a representative cluster, SLIC uses a distance function that considers both the data value similarity as well as their spatial proximity. In this work, we used the distance function as suggested in~\cite{dutta_slic_pvis,slic_ucdavis}:
\begin{equation}
D(x,y) = \gamma \cdot \vert \vert c_x - p_y \vert \vert_2 + (1-\gamma)\cdot \vert v_x - v_y \vert \label{slic_distance}
\end{equation} 
Here, $c_x$ is the location of the cluster center $x$ and $p_y$ is the location of data point $y$. $v_x$ and $v_y$ are the scalar values at xth cluster center and yth data point respectively. The value of $\gamma$ ($ 0 <= \gamma <= 1$, and $\gamma + (1-\gamma) = 1$) is chosen based on the requirement to specify weightage for spatial vs value components. We have set $\gamma=0.3$ for this work to assign higher weightage on data values. Such a distance function ensures that the produced supervoxels are spatially contiguous and as homogeneous as possible. Note that since the global density field is constructed in the root processing node by combining contributions from all other processing units, we compute SLIC partitioning only in the root node using the global density field.  

Figure~\ref{fig:insitu_transformations}c shows the result of SLIC algorithm  applied to the density field in Figure~\ref{fig:insitu_transformations}b. In this example, SLIC is performed in image space for illustration. In in situ, SLIC is performed on the 3D density field. For a more detailed description of SLIC, readers are referred to~\cite{slic,dutta_slic_pvis}. After SLIC, density values of data points in each SLIC partition are modeled using a statistical distribution to construct the distribution field, resulting in a compact representation.
Since SLIC produces a homogeneous partitioning, a single Gaussian distribution for each partition is used to model the density values as shown in Figure~\ref{fig:insitu_transformations}d. The use of a single Gaussian was shown to be sufficient and storage efficient by Dutta et al.~\cite{dutta_slic_pvis} when a homogeneous partitioning scheme such as SLIC was used. Each SLIC cluster is modeled using a Gaussian and by doing so, the density field is converted to a distribution field.

\subsection{Bubble Similarity Field (BSF) Creation}
The SLIC-based distribution field is further classified to produce bubble similarity fields (BSF) where high similarity values indicate a higher chance of being part of a bubble. 
We estimate the density field and collect density values from the user highlighted region (Section~\ref{feature_selection}), shown in Figure~\ref{fig:feature_select}, to define the Gaussian distribution used as the statistical signature of a bubble in the analysis pipeline. During in situ processing, we compute the statistical similarity of each cluster's Gaussian distribution to that of the user-provided signature Gaussian. By doing so, each cluster gets a statistical similarity score. Finally, a new scalar field, called Bubble Similarity Field (BSF), is constructed where each point indicates the degree of statistical similarity to that of the signature bubble distribution. The Bhattacharyya distance~\cite{bhattacharya_dist_gaussian} is used to measure the similarity between two Gaussian distributions. Since the Bhattacharyya distance provides a closed-form solution and can be computed efficiently, it is found to be suitable for estimating the distance between Gaussians in the in situ environment. The Bhattacharyya distance, $D_{Bh}(g_1,g_2)$, between two Gaussian kernels is defined as: 
\begin{equation} \label{eq:bhatta_dist_Gaussian}
{\scriptstyle
D_{Bh}(g_1,g_2) = \frac{1}{8}(\mu_1 - \mu_2)^{\mathsmaller T}\Bigg(\frac{\sigma_1 + \sigma_2}{2}\Bigg)^{\mathsmaller -1}(\mu_1 - \mu_2)^{\mathsmaller T} + \frac{1}{2}ln\Bigg[\frac{\lvert\frac{{\sigma_1 + \sigma_2}}{2}\rvert}{\sqrt{\lvert\sigma_1\rvert\lvert\sigma_2\rvert}}\Bigg]
}
\end{equation}
where $\mu_1$, $\mu_2$ and $\sigma_1$, $\sigma_2$ are the mean and standard deviation of the Gaussian kernels $g_1$, $g_2$ respectively. A lower value of $D_{Bh}(g_1,g_2)$ indicates a higher degree of similarity.  Before storing  values in the similarity field, we normalize the $D_{Bh}(g_1,g_2)$ values and subtract it from 1 to be able to interpret them as similarity scores in post hoc analysis phase. Hence, the values of the BSF range from 0-1 where high values indicate higher statistical similarity to the target feature. Again note that the BSF is computed in the root processing node only and this step does not need any additional data communication. Visualization of a BSF is provided in Figure~\ref{fig:insitu_transformations}e. It can be seen that the low particle density regions in Figure~\ref{fig:insitu_transformations}a are classified as the bubbles with high similarity values (the blue regions). 

\subsection{Particle Rise Velocity Field (PVF) Generation}
While studying bubble dynamics, domain experts are also interested in the particle velocity profile around each bubble as they rise through the fluidized bed. Therefore we also estimate a second scalar field which summarizes the vertical rise velocity (x-direction) component of the particle velocity field. This second scalar field is stored along with the BSF to enable flexible bubble dynamics analysis. The estimation of particle rise velocity scalar field (PVF) follows similar steps that were used for creating the 3D spatial histogram and is again distributed. In this case, the average rise velocity (x-direction) of the particles for each bin is estimated. First, the cumulative rise velocity of all the particles is computed for each bin locally, and then, using a parallel reduction operation, the global velocity histogram is computed. Finally, the average rise velocity per bin is computed by dividing the global 3D velocity histogram by the global density histogram, which is already estimated for computing the particle density field. The particle velocity histogram is then converted to a scalar field following similar steps used for the generation of the density field from the density histogram. A visualization of PVF for the raw particle data, shown in Figure~\ref{fig:insitu_transformations}a, is provided in Figure~\ref{fig:insitu_transformations}f. The red and yellowish regions in this image indicate particles with positive rise velocities, i.e., the particles in those regions are rising upward, and the blue regions contain particles that are moving downward. At the end of in situ processing, these two scalar fields are stored to disk for each time step.

\section{Post Hoc Bubble Dynamics Analysis}

\subsection{Bubble Extraction and Characterization}
\begin{figure}[tb] 
\centering
\includegraphics[width=\linewidth]{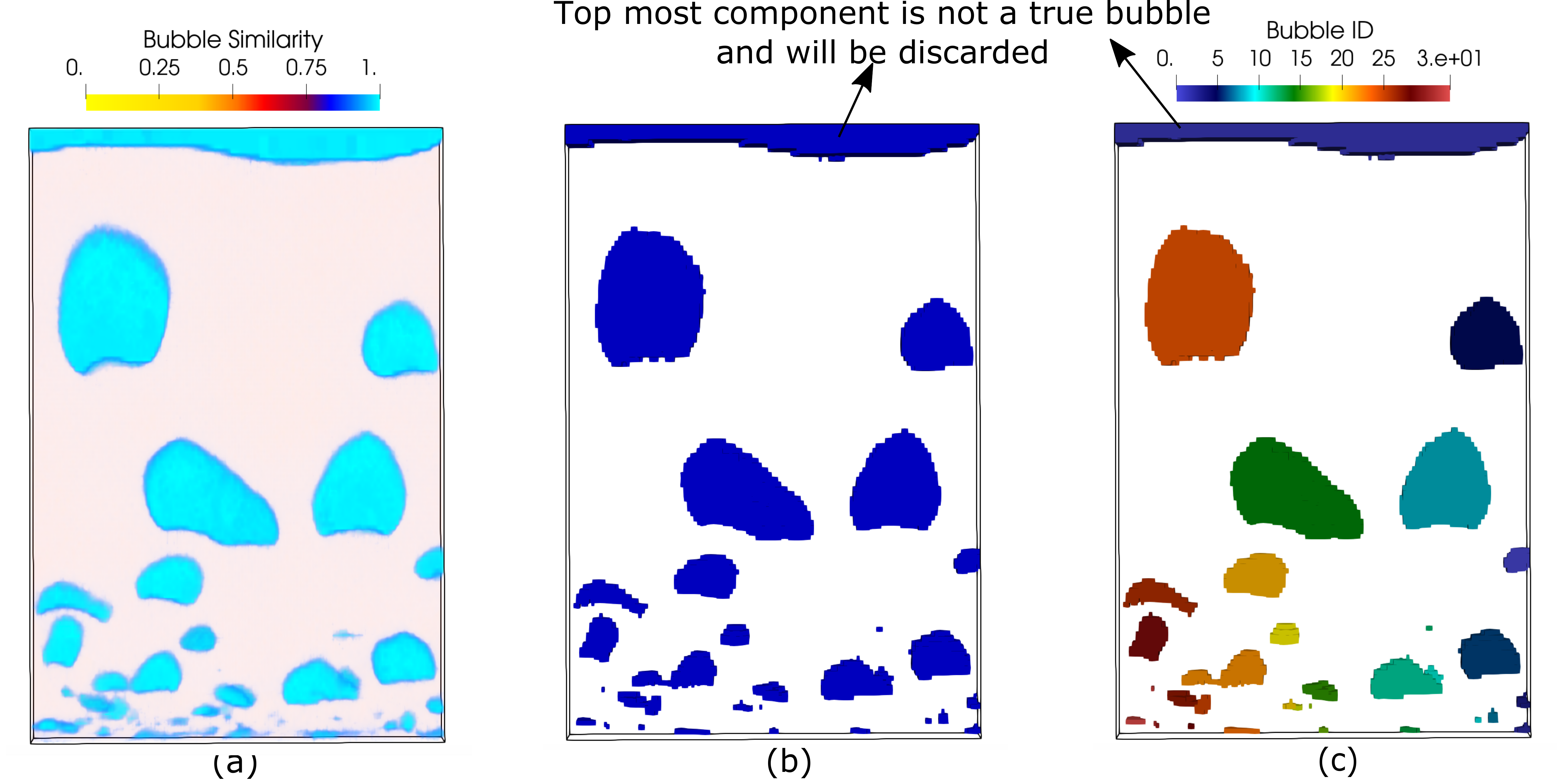}
\caption{(a) Visualization of bubble similarity field, (b) Segmented bubbles, (c) Isolated bubbles when connected components are detected.}
\label{fig:BSF_segmentation}
\end{figure}

To study bubble details, the first step is to extract individual bubbles from the BSFs. Similarity values in BSF range from 0 to 1 and higher valued regions indicate higher possibility of being part of a bubble. Experts can inspect several BSFs interactively at the beginning of analysis and specify a desired degree of similarity value to segment the BSFs. This similarity threshold value is applied to all BSFs from all time steps to produce consistent segmentation results. Next, a connected component (CC) algorithm is applied to the segmented results and each component is treated as a bubble feature. Figure~\ref{fig:BSF_segmentation} shows results of this bubble extraction process for a time step. Figure~\ref{fig:BSF_segmentation}a shows the BSF with blue regions indicating different bubbles. Figure~\ref{fig:BSF_segmentation}b depicts the results when segmentation is performed for a similarity threshold value of $0.92$. Finally, results of the connected component algorithm are shown in Figure~\ref{fig:BSF_segmentation}c where different bubbles are colored by their unique component ids. Note that the simulation data often produces a void region on the top of the fluidized bed and has very low particle density. Although it is also detected as one of the potential bubbles (marked in Figure~\ref{fig:BSF_segmentation}b and Figure~\ref{fig:BSF_segmentation}c), these void regions are not considered true bubbles.
Once all bubbles are extracted, several characteristics for each bubble feature are computed. The salient characteristics~\cite{Boyce_single_bubble,BubbleTree} are considered volume, centroid, and aspect ratio of each bubble. The approach  in~\cite{Samtaney1994} was used for estimating bubble volume and centroid. For aspect ratio, we used the measure provided in~\cite{Boyce_single_bubble}. 
\begin{figure*}[tb] 
\centering
\frame{\includegraphics[height=2.8in, width=0.8\linewidth]{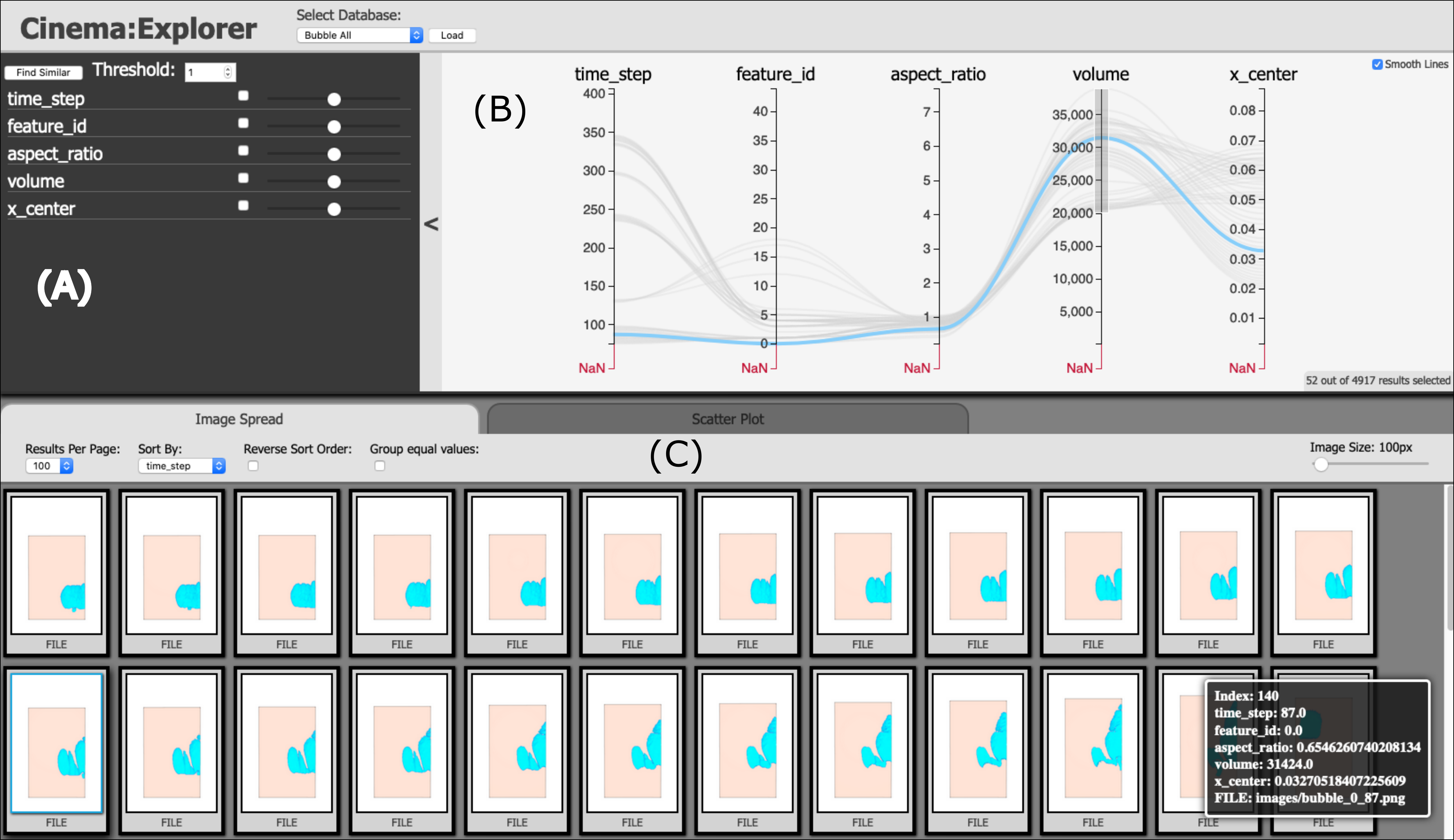}}
\caption{Visualization of overall bubble characteristics using CinemaExplorer. Bubbles with large volumes are queried in the PCP and the filtered bubbles are displayed in the image view at the bottom. The PCP shows other relevant bubble characteristics such as aspect ratio, position of bubbles in the rising direction, time step, etc.}
\label{fig:bubble_all_example}
\end{figure*}
\subsection{Bubble Tracking} \label{tracking_subsection}
Apart from isolating and characterizing each bubble using its salient characteristics, another critical requirement is to be able to analyze the temporal dynamics of the bubbles on demand. To achieve this, we modified a well-known overlap-based feature tracking algorithm~\cite{Silver96volumetracking} by adding an estimation of feature matching confidence to it. In the volume tracking algorithm presented in~\cite{Silver96volumetracking}, the feature correspondence problem was solved by finding overlapping objects in consecutive time steps. This technique assumes the availability of sufficient temporal resolution of data so that overlapping-based correspondence can correctly identify the features over time. However, there exists a minimal chance of incorrect correspondence. To address this issue, we have used the Dice similarity index~\cite{dice_sim} to detect bubble overlapping in consecutive time steps because of its robustness over other similar methods~\cite{zou04}. Dice index allows estimation of overlap between two sets with a minimum of $0$ indicating no overlap, and a maximum value of $1$ for complete overlap. Formally given two sets of points $A$ and $B$, their Dice similarity index $DI(A, B)$ is measured as:  
\begin{equation}
 DI(A,B) =  \frac{2\left| A \cap B \right|}{\left| A \right| + \left| B \right|}    \label{eq:dice_sim}
\end{equation}
The use of the Dice index allows us to detect the overlap between the two segmented bubbles in 3D and also quantify a similarity score reflecting the degree of matching. The Dice index value can be interpreted as matching confidence and when an abruptly low value is found, such a time step is flagged for further investigation. In our visualization tool, values of the Dice indices are presented
so experts can make informed judgments on the tracking results.

The analysis method enables tracking of a single bubble as well as all the bubbles from a selected time step. While the single bubble tracking allows the experts to focus on the evolution of a specific feature of interest, the collective tracking of all bubbles from an interesting time step allows opportunities for comparative visualization where collective bubble dynamics can be studied. We track all the bubbles both forward and backward in time as long as we can find a corresponding bubble that overlaps with the current bubble selected, capturing the life cycle of a bubble from birth to death. In bubble evolution, it is important to study the different evolutionary events that the bubbles go through such as splitting and merging with neighboring bubbles. In a typical fluidized bed simulation, small bubbles form at the bottom of the bed, merging and splitting as they rise through the fluidization zone before bursting out at the top.
In our method, we track bubble volumes to explore potential merge and split events. During tracking, a sudden rise in the bubble volume indicates a merge event while a significant drop in volume would indicate splitting~\cite{dutta_vis15}. From the tracking results, we also estimate the bubble rise velocity, an important bubble characteristic of interest to the domain experts. The bubble rise velocity is computed from differences of the centroids of matched bubbles from the consecutive time steps as suggested in~\cite{BubbleTree}.

\section{Visual Exploration of Bubble Dynamics}
We present bubble dynamics through several interactive visualization interfaces such that the experts can query, filter, and track bubbles to understand their interaction in detail. The custom visualization interfaces used are developed under the Cinema project~\cite{Jim_SC_cinema,cinema_url}, which provides innovative solutions for capturing, storing, and exploring large-scale data sets using an image-based approach. The Cinema project offers several open-source visualization tools targeted for specific visualization tasks. We have customized two Cinema viewers, CinemaExplorer \cite{cinema_explorer} and CinemaView \cite{cinema_view}, to visualize bubble dynamics. 

In our fluidized bed simulations, the Y dimension is typically small compared to the other two dimensions (X and Z), so the experts during their exploration visualize the bubbles from the X-Z plane, which maximizes the viewing area of the simulation domain. Therefore, after bubbles are extracted, we use volume rendering to render each bubble viewed from the the X-Z plane and store visualizations according to Cinema specifications~\cite{cinema_url}. First we visualize the field in ParaView and generate a Python-based ParaView rendering script containing the rendering specifications and then run the script to produce visualizations of all the bubbles in batch mode and store the results as images in standard PNG format. All rendering parameters (viewing and camera parameters, and transfer functions for volume rendering) are fixed for all the bubbles. Note that this operation is done once and takes only $\sim$5 minutes to produce all the images. This results in an image-based database that includes bubble characteristics and a visual representation of the bubbles. The Cinema database is then explored interactively using the Cinema viewers. The intuitive and well-known visualization techniques used in our Cinema viewer, such as Parallel Coordinates Plots (PCP) and Scatterplots with brushing and linking capabilities, are used to present visualization results to the domain experts. Along with these interactive tools, we have also developed a visualization tool that allows visualization of bubble tracking directly in the three dimensional domain for the experts to study bubble interaction in 3D.

\begin{figure}[tb] 
\centering
\frame{\includegraphics[width=\linewidth]{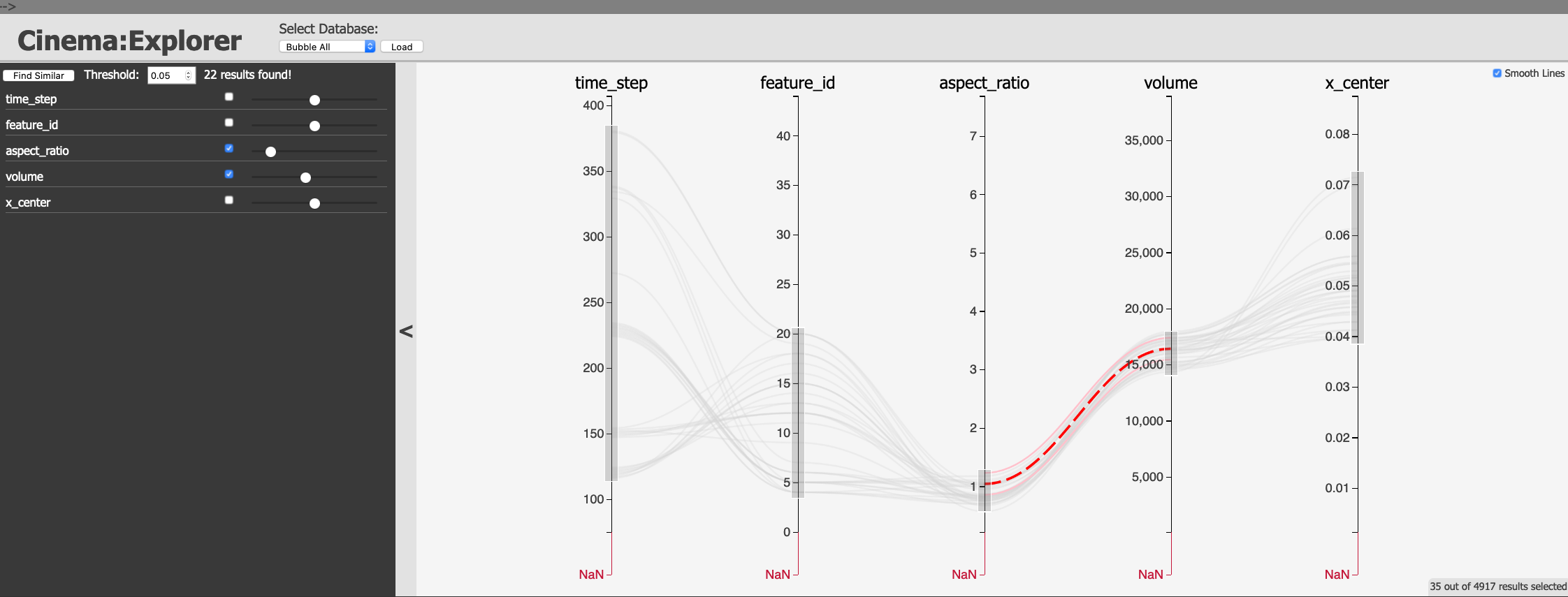}}
\caption{Interactive query of multiple bubble characteristics using the query panel of CinemaExplorer. Bubbles that fall within a range of aspect ratio and volume are queried in this example.}
\label{fig:PCP_query_illustration}
\end{figure}
\subsection{Overview Visualization of Bubble Characteristics}
\begin{figure}[tb] 
\centering
\frame{\includegraphics[width=0.9\linewidth]{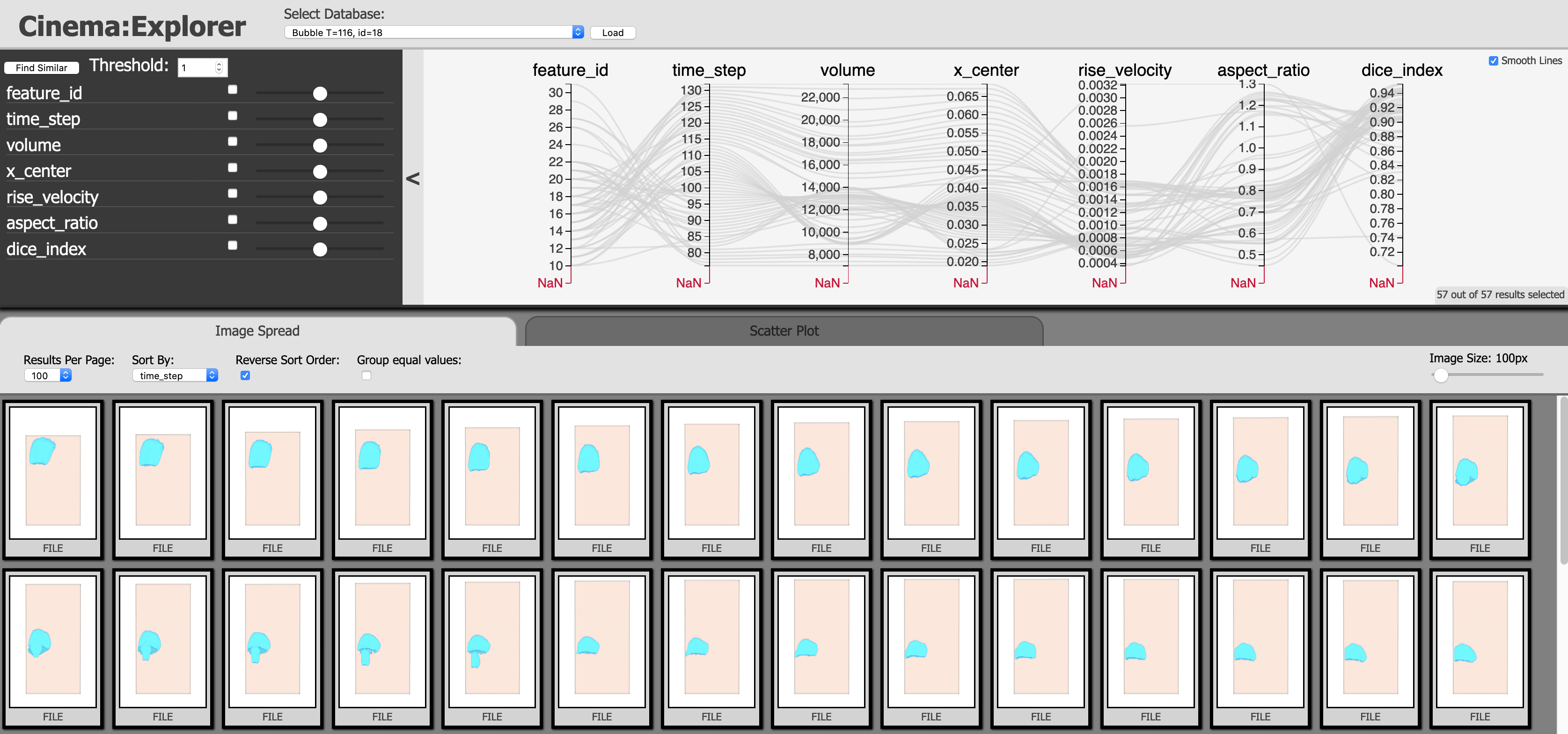}}
\caption{Visual exploration of the results of a single bubble tracking using CinemaExplorer. The initial feature was selected from time step = 116 with feature id = 18.}
\label{fig:single_track_example}
\end{figure}

\begin{figure}[tb] 
\centering
\includegraphics[width=0.9\linewidth]{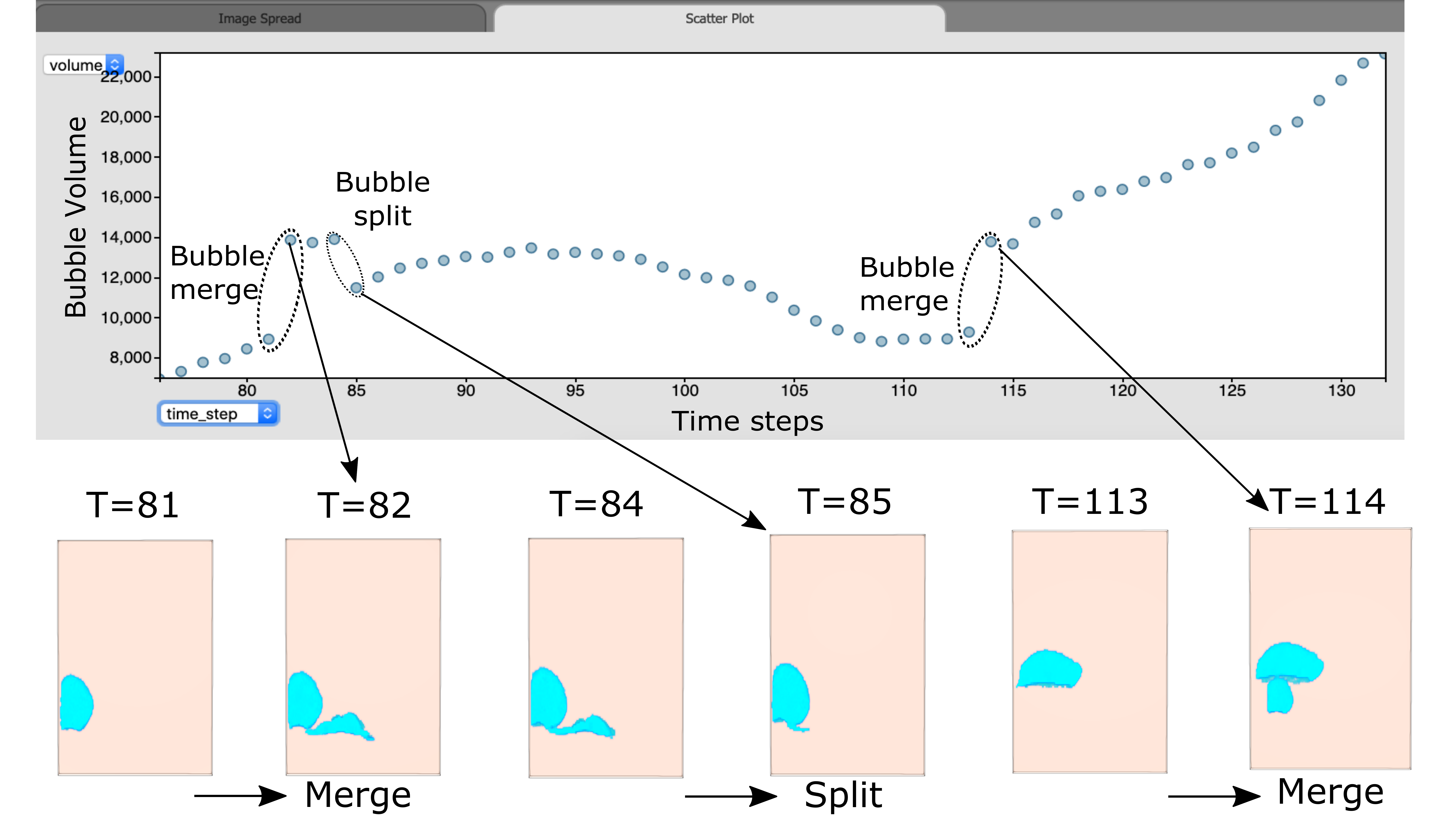}
\caption{(Top) Volume vs time scatterplot revealing bubble merge and split events with the initial bubble selected at time step=116 and feature id=18. (Bottom) Detected events are shown for different time steps.}
\label{fig:combined_116_18}
\end{figure}

The first step in our visual analysis pipeline uses CinemaExplorer to study overall bubble characteristics so that the experts can obtain a general understanding. A typical CinemaExplorer viewer has three panels as highlighted in Figure~\ref{fig:bubble_all_example}: (A) the query panel, (B) the Parallel Coordinates Plots (PCP) panel, and (C) image spread panel. The query panel (A) provides multivariate query capabilities as shown in Figure~\ref{fig:PCP_query_illustration}. Using a slider provided for each bubble characteristic, the user can select a value range for specific attribute to filter bubbles. In Figure~\ref{fig:PCP_query_illustration}, data is filtered using aspect ratio and volume attributes. The red dotted line shows the central value around which the filtering is done, and the dotted red lines show the filtering range. Panel (B) shows the PCP where the bubble characteristics are represented as parallel axes and the characteristics of each bubble are represented by a polyline. The PCP supports interactive brushing and axis reordering. The visualization of the filtered bubbles is shown in the image spread view, Panel~(C). Image size can be adjusted using a slider on the top right corner of the panel. On hovering the mouse over the bubble images, a pop-up appears displaying its attribute values, and the corresponding polyline in the PCP is highlighted in blue (Figure~\ref{fig:single_track_example}). Filtering in PCP or query panel automatically updates results in the image view. 
An additional tab provides a scatterplot visualization of any two selected bubble characteristics (see, e.g., Figure~\ref{fig:combined_116_18}). Such scatterplots are found to be effective in displaying the temporal evolution of bubble characteristics. Using these visualizations, the experts can identify bubbles based on their aspect ratio, volume, position, etc., for further investigation.

\subsection{Visualization of Bubble Tracking Results}

\subsubsection{Visualization Using the Cinema-based Viewer} 
Once users identify an interesting bubble, it is tracked over time and the analysis result is presented using CinemaExplorer as shown in Figure~\ref{fig:single_track_example}. In this example, a relatively large bubble was selected from time step $116$ and tracked both forward and backward in time. Note that the PCP has two additional axes showing information about the rise velocity and the values of the Dice index estimated during tracking. The Dice index values are mostly high, indicating stable tracking results. Using this visualization, the experts can interactively explore the life cycle of the selected bubble along with all potential merge and split events associated with it. The CinemaExplorer image spread panel presents the visualization of the bubbles side-by-side. Showing the tracking results in such a way allows the experts to easily compare the evolution of a bubble within neighboring time steps without losing the mental map. As discussed above, users can also filter results based on the bubble attribute values by using both the PCP and the query panel. The scatterplot functionality in the CinemaExplorer viewer is useful to investigate the details of the various evolutionary events such as merging and splitting. In Figure~\ref{fig:combined_116_18}, the scatterplot of bubble volume vs time steps is shown at the top. In the simulation, bubbles are generally formed at the bottom of the bed and rise upward, moving through the fluidized bed. Time steps when bubble volumes abruptly change are easily identified in the scatterplot. Several such events are highlighted in the plot with the specific events that caused the change in bubble volume. The visualization of these events for this bubble is provided at the bottom of Figure~\ref{fig:combined_116_18}. It can be seen that, indeed, the abrupt increase in bubble volume indicates merge events and split events correspond to a sudden decrease in bubble volume. 

\begin{figure}[tb] 
\centering
\includegraphics[width=\linewidth]{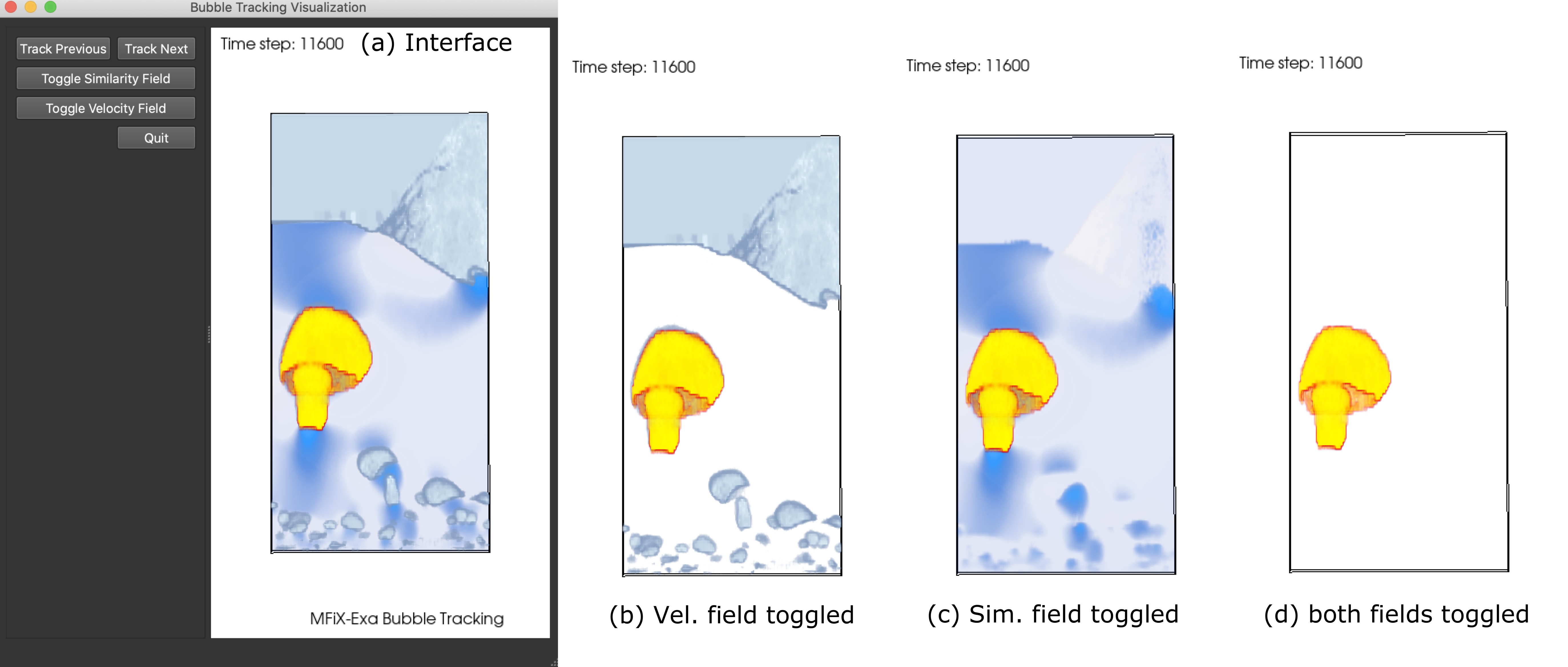}
\caption{Visualizing bubble tracking directly in 3D domain using our feature tracking interface shown in Figure~\ref{fig:track_3D}(a). The interface shows the tracked bubble (highlighted in yellow). All the other bubbles are shown as context and, in the background, the regions that have high particle rise velocity is also presented (colored in light blue). The interactive interface allows the users to turn on and off different visualization components. In Figure~\ref{fig:track_3D}(b) PVF is turned off, in Figure~\ref{fig:track_3D}(c) the bubbles are turned off, and Figure~\ref{fig:track_3D}(d) depicts the visualization of only the tracked bubble.}
\label{fig:track_3D}
\end{figure}

\subsubsection{Visualizing Bubble Tracking in 3D Spatial Domain}
\bmark{In the previous section, we have described how we use an image-based approach to facilitate the interactive study of bubble dynamics. Many characteristics of the tracked bubbles can be analyzed using the CinemaExplorer tool. However, to allow the application scientists to explore the 3D nature and shape of the bubbles, we have developed a  feature tracking visualization tool that uses 3D volume rendering techniques to provide interactive visualization of bubbles. This tracking interface is shown in Figure~\ref{fig:track_3D}(a). The interface is developed in Python and integrates VTK~\cite{VTK} and QT for rendering and interaction capabilities. The control panel on the left provides means of interaction with the results. The users can interactively change time steps. This interface provides an integrated visualization of bubble tracking results where the tracked bubble is highlighted in yellow. In the background, all the other bubbles are shown with grey color for context. Finally, the regions where the PVF have high values, i.e, the regions where the particles are moving faster than others are shown using volume rendering and are colored in light blue as seen from  Figure~\ref{fig:track_3D}(a). Different visualization components convey different information and the combined view provides a holistic understanding for the users who can study the temporal bubble evolution while also visualizing the dynamic nature of the particle velocities around the bubbles. At any time, the users can interactively turn on or off the additional bubbles or the PVF visualization to focus on the tracked bubble. In Figure~\ref{fig:track_3D}(b), we show the rendering window when PVF is turned off, in Figure~\ref{fig:track_3D}(c) the additional bubbles are turned off, and finally in Figure~\ref{fig:track_3D}(d) only the tracked bubble is shown. Users can switch between these different visualization modes using the control panel.}

\bmark{
\subsubsection{Scatterplot Matrices for Correlation Study} 
Besides visualizing the bubbles through the CinemaExplorer tool and our 3D tracking interface, the users are also interested in discovering correlations among various bubble characteristics. To do this, we provide a scatterplot matrix, (SPLOM)~\cite{splom}, visualization for the users where all pair scatterplots are shown simultaneously in a matrix-based layout (as shown in Figure~\ref{fig:splom_combined_154}). SPLOM gives a quick overview visualization where the users can inspect correlations between any two bubble characteristics. Once they find an interesting pair, they can use the scatterplot in CinemaExplorer for a detailed study where the users can interactively highlight points in scattaerplot and inspect the corresponding bubble characteristics values. 
Therefore, by combining CinemaExplorer functionality, our feature tracking interface, and SPLOMs, we are able to present in depth information about the temporal bubble dynamics to the application experts. } 

\subsection{Comparative Visualization of Bubble Dynamics}
\begin{figure}[!tb] 
\centering
\frame{\includegraphics[width=\linewidth]{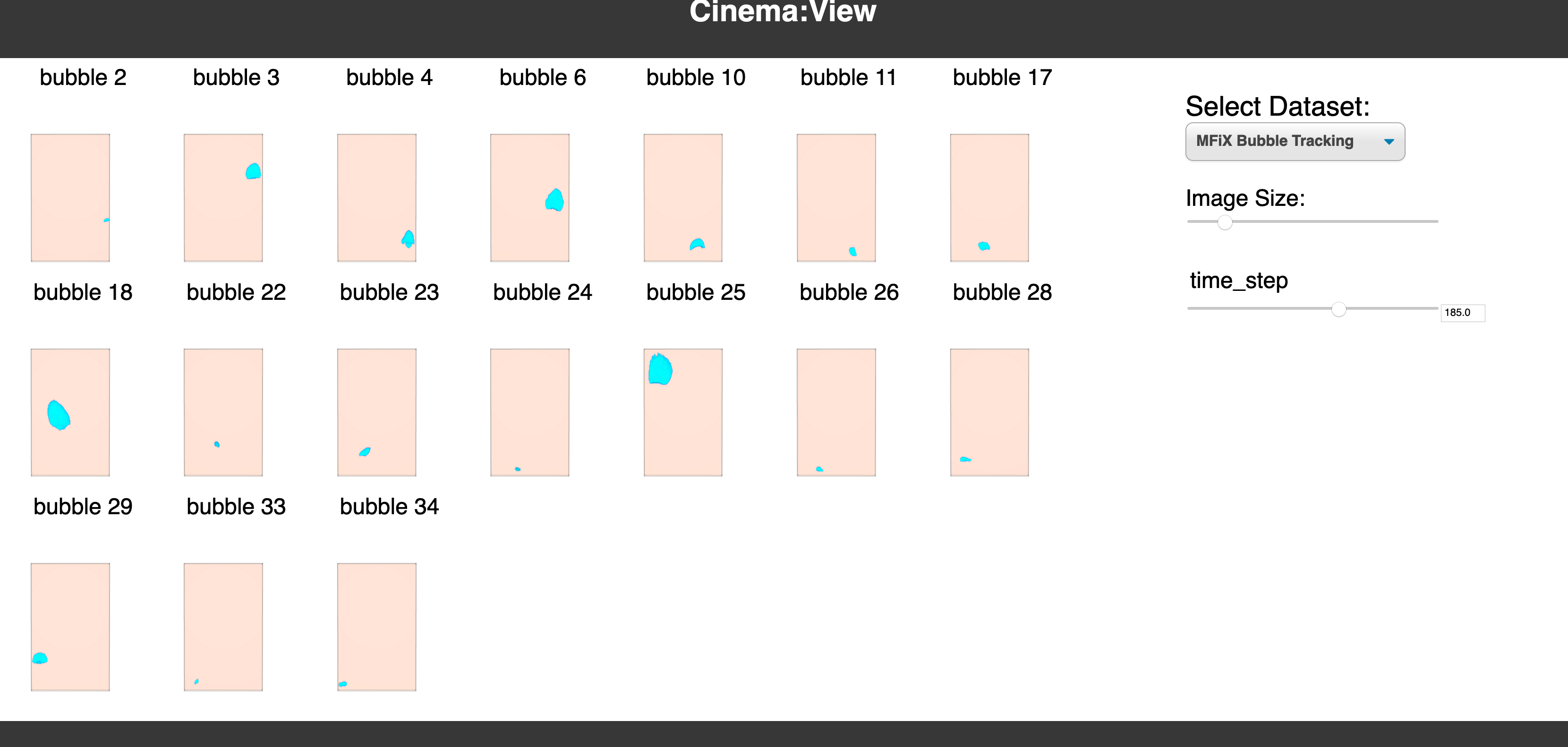}}
\caption{Comparative visualization of tracking of all bubbles from a time step using CinemaView tool.}
\label{fig:collective_track_example}
\end{figure}

The proposed visual analytic system also allows collective tracking and visualization of all the bubbles from a specific time step. Collective bubble tracking enables comparative visualization among all the bubbles, showing their evolution over time. To visualize the tracking results, we use CinemaView as shown in Figure~\ref{fig:collective_track_example}. CinemaView provides comparative visualization among multiple bubble features, where the tracking of each individual bubble is represented as a separate Cinema database. CinemaView can load multiple Cinema databases and allows exploration via a common attribute, in this case, time steps. The size of the images for each bubble can be adjusted as necessary and, using the time step slider, the experts can interactively study how bubbles evolve over time. Over the course of the time window, if some bubbles dissipate, then that image is replaced with an empty window indicating bubble dissipation event. Similarly, when an empty window is replaced with a bubble image forward in time, a bubble birth is observed.

\section{Results, Discussion, and Expert Feedback}
We validate the above findings on bubble dynamics through expert feedback and by verifying our results with findings in  multiphase flow research literature. We provide details of bubble evolution from the tracking results and explore the relationships among various bubble characteristics. Note that, we have linearized the time step numbers during analysis. The actual simulation time step numbers can be found by multiplying $100$ with the numbers reported here.

\subsection{Analysis of Bubble Characteristics}
\begin{figure}[tb] 
\centering
\includegraphics[width=0.9\linewidth]{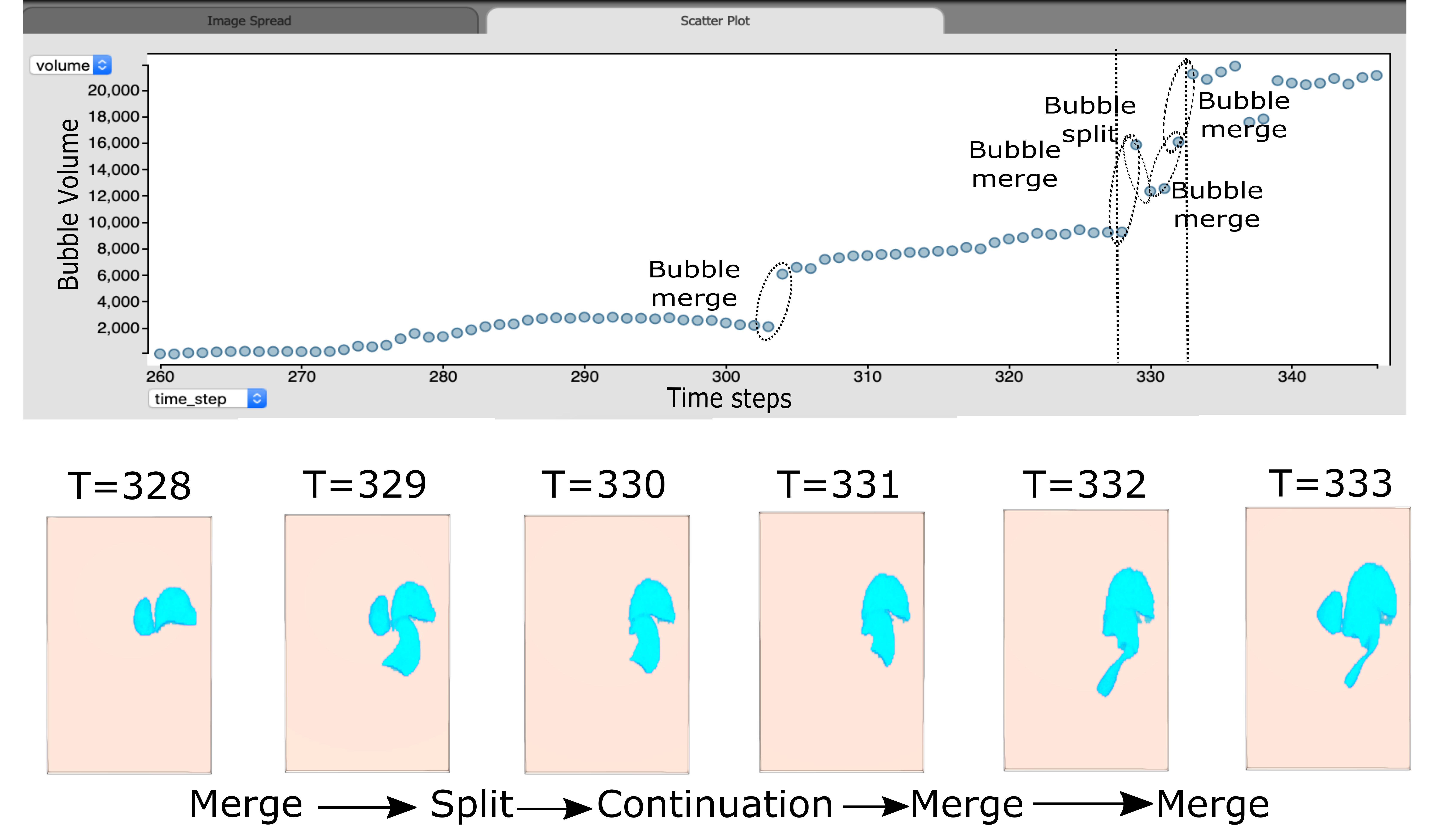}
\caption{(Top) Volume vs time step Scatterplot revealing bubble merge and split events. The initial feature was selected from time step = 338 with feature id = 7. (Bottom) Visualization of merge and split events for this bubble. Detected merge and split events are shown for a time window indicated by vertical dotted lines in the Scatterplot.}
\label{fig:combined_338_7}
\end{figure}

Bubbles in the fluidized bed simulation form consistently at the bottom of the bed and rise upward before finally reaching the bed surface. Initially bubbles are generally small, and gradually grow as they rise, merging with other bubbles. After studying bubble tracking results for multiple bubbles, we observed that bubble merge events are more frequent than bubble splitting and as a result, the majority of the bubbles grow in size over time before reaching the top. This observation was confirmed by our domain expert. It is also noted that while studying bubble tracking results using CinemaExplorer, the scatterplot between bubble volume and time steps effectively shows this consistent trend for the majority of the bubbles studied. Figure~\ref{fig:combined_338_7} illustrates how bubble volume changes over time with sudden changes signifying merge/split events. Also, during investigation of these events, by filtering the bubbles using aspect ratio (width/height) values, users can easily locate bubbles which are spherical in shape (aspect ratio $\approx$ 1). Such bubbles are often of interest to the experts~\cite{Boyce_single_bubble}. We further observe that the bubble rise velocity increases slowly as the bubble volume increases. Consistent with existing literature, bubble rise velocity remains relatively constant if bubble volume remains constant~\cite{Boyce_single_bubble}, whereas rise velocity increases with increasing volume~\cite{Boyce_single_bubble,Davies1988TheMO}.

\begin{figure}[tb] 
\centering
\includegraphics[width=\linewidth]{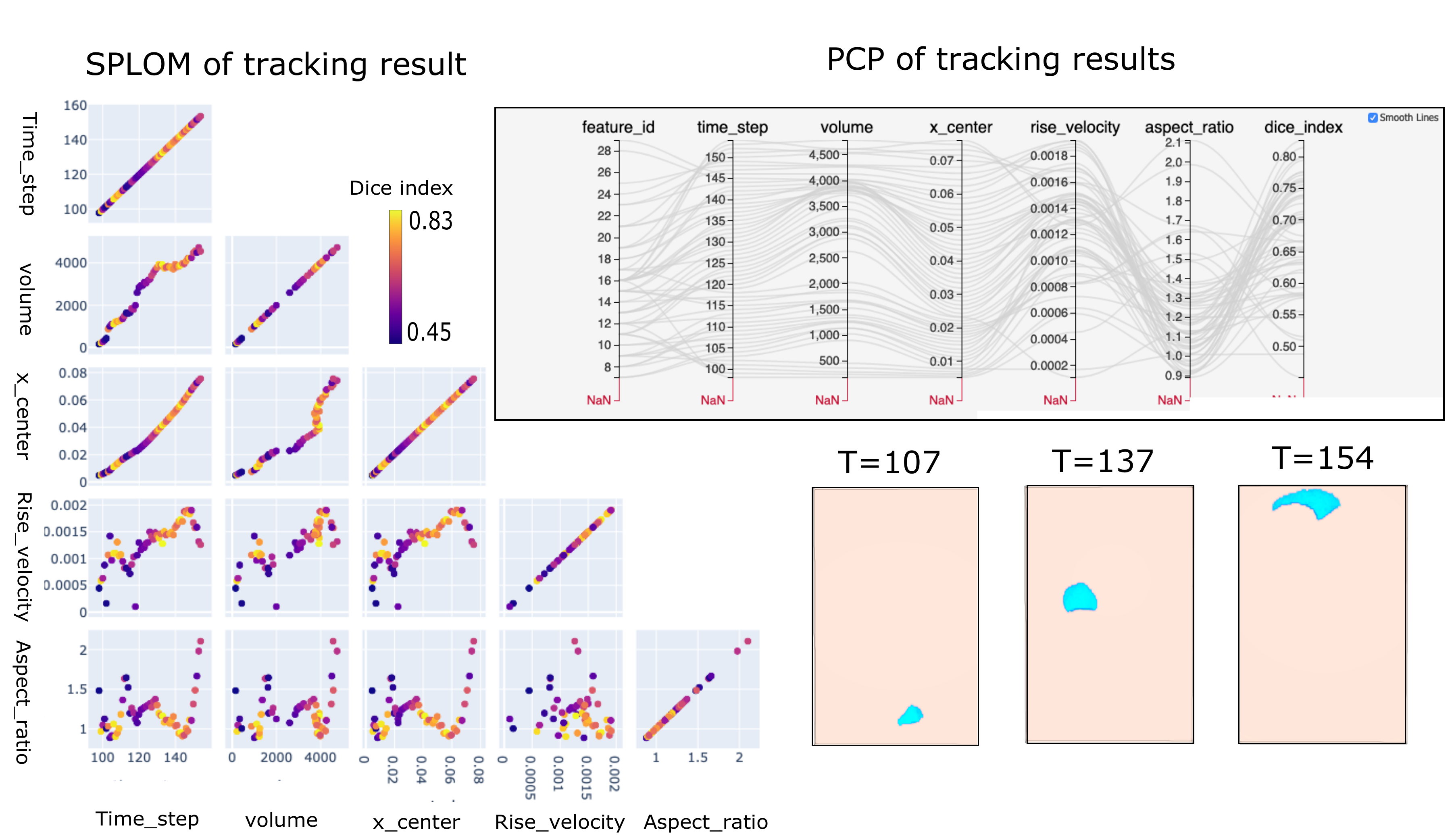}
\caption{Visualization of SPLOM generated for studying relationships among various salient bubble characteristics for a bubble tracked over time. The bubble was selected from time step = 154 with feature id = 16. The PCP from the CinemaExplorer is also shown. At the bottom right, visualizations of the selected bubble from three different time steps are provided. It can be seen that the bubble increases in size over time as it evolves through the fluidized bed.}
\label{fig:splom_combined_154}
\end{figure}

\textbf{SPLOM-based correlation study.} 
 In Figure~\ref{fig:splom_combined_154}, a SPLOM is shown for a specific bubble tracked over time. The SPLOM is plotted as a lower triangular matrix where each cell is a scatterplot between two specific bubble characteristics. The points in the SPLOM scatterplots are colored by the Dice similarity index, indicating the bubble matching confidence during tracking at each time step. For comparative analysis, we also provide the PCP from CinemaExplorer and three representative visualizations of the bubble, one from an initial time step (T=107), the second from an intermediate time step (T=137), and the third from a later time step (T=154). By observing the SPLOM, we find that the bubble volume and rise velocity increases with time (Cell [1,0], and [3,0] in SPLOM). Furthermore, the rise velocity and bubble volume (Cell [3,1]) is also correlated directly. Similar relationships among bubble volume and rise velocity were documented in previous studies~\cite{Boyce_single_bubble,Davies1988TheMO}. This correlation between bubble volume, rise velocity, and time is also observed in the PCP in Figure~\ref{fig:splom_combined_154} among PCP axes time step, volume, and rise velocity respectively. Note that the axis x\_center indicates the position of the bubble in the rising direction and since the bubble rises gradually over time, x\_center is also found to be correlated with volume and rise velocity as seen from both SPLOM and PCP.

\subsection{Particle Dynamics Exploration}
\begin{figure}[tb] 
\centering
\includegraphics[width=\linewidth]{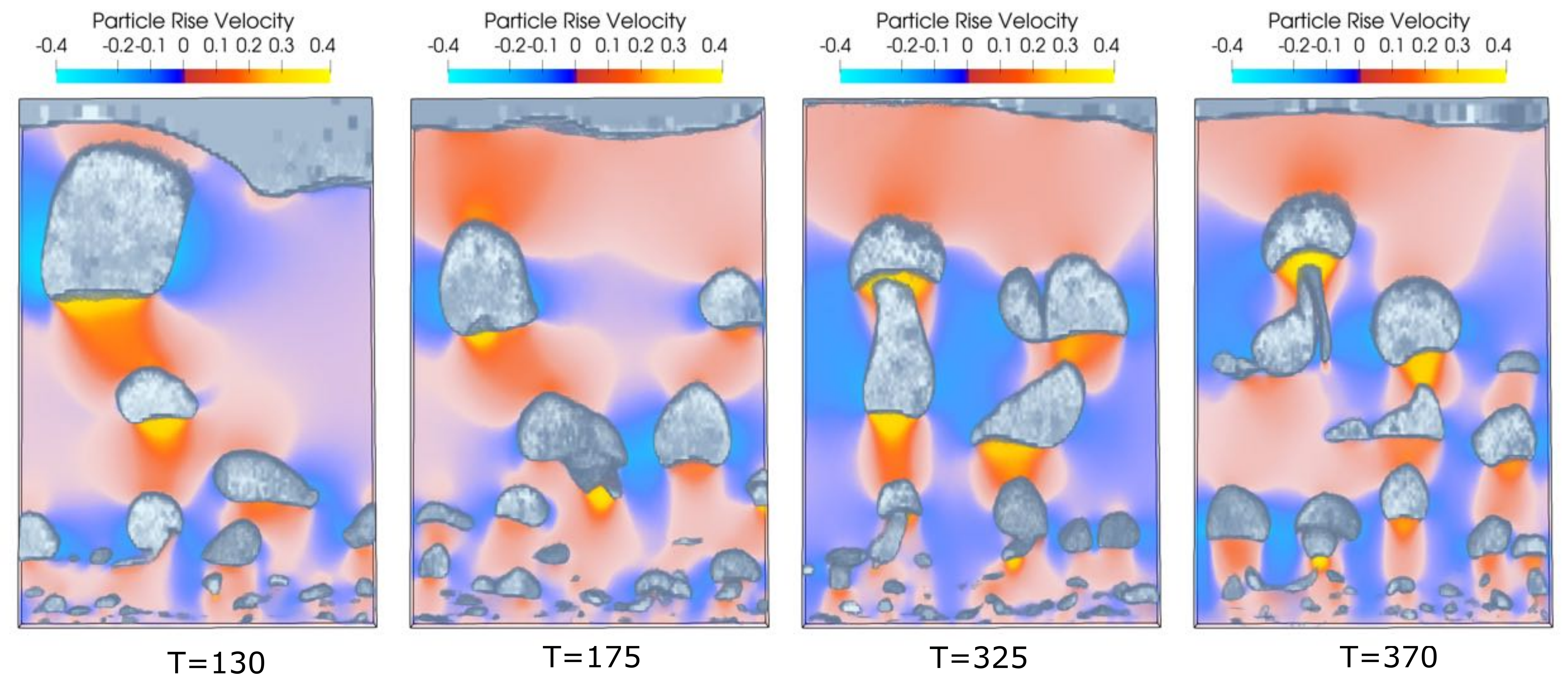}
\caption{Visualization of particle rise velocity fields (PVF) for different time steps. The blue regions indicate particles with negative velocity (downward movement)
and the red to yellow regions show particles with positive rise velocity(upward movement). The PVFs are computed in situ from the raw particle velocity fields and can be used to effectively study the dynamics of particles around bubbles.}
\label{fig:pvf_viz}
\end{figure}

The study of particle dynamics around the bubbles is an important task for the domain scientists as this allows exploration of the hydrodynamics in the fluidized bed~\cite{Boyce_single_bubble}. Visualization of particle rise velocity fields (PVF), computed in situ,  complement the above analyses, helping to explain particle dynamics around the bubbles as they form, rise, and merge. Figure~\ref{fig:pvf_viz} shows a visualization of PVF for four different time steps representing different states of the simulation. The selected colormap shades particles moving upward (positive velocity) red-to-yellow whereas downward moving particles are blue. An important observation is that particles around the bubbles move downward, consistent with the literature~\cite{Boyce_single_bubble}. The domain expert explains that this happens so that the particles around the bubble can move down and replenish those carried upward in the wake of the bubbles. The wake of a bubble is the region immediately behind it (the red-to-yellow region in the images). Similarly, it is observed that particle velocities above and below bubbles are high. The velocity is higher underneath bubbles in the wake than above bubbles which is seen from the yellow regions at the bottom of bubbles. This distribution of low-velocity particles around the bubbles and high-velocity particles on the top and bottom of the bubbles generates a circular flow causing the bubbles to rise before breaking through the freeboard~\cite{Boyce_single_bubble}. 

\begin{figure}[tb] 
\centering
\includegraphics[width=0.9\linewidth]{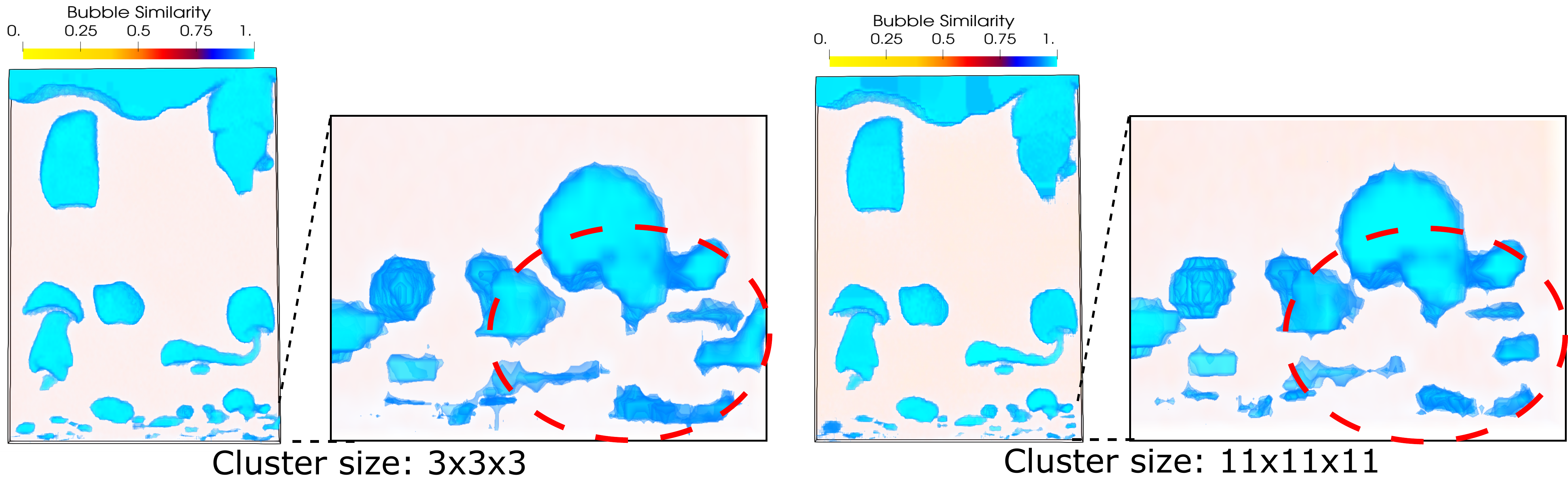}
\caption{Impact of different cluster sizes on bubble detection. For larger cluster sizes, small bubbles might be detected inaccurately.}
\label{fig:slic_cluster_param}
\end{figure}

\begin{figure}[tb] 
\centering
\includegraphics[width=0.7\linewidth]{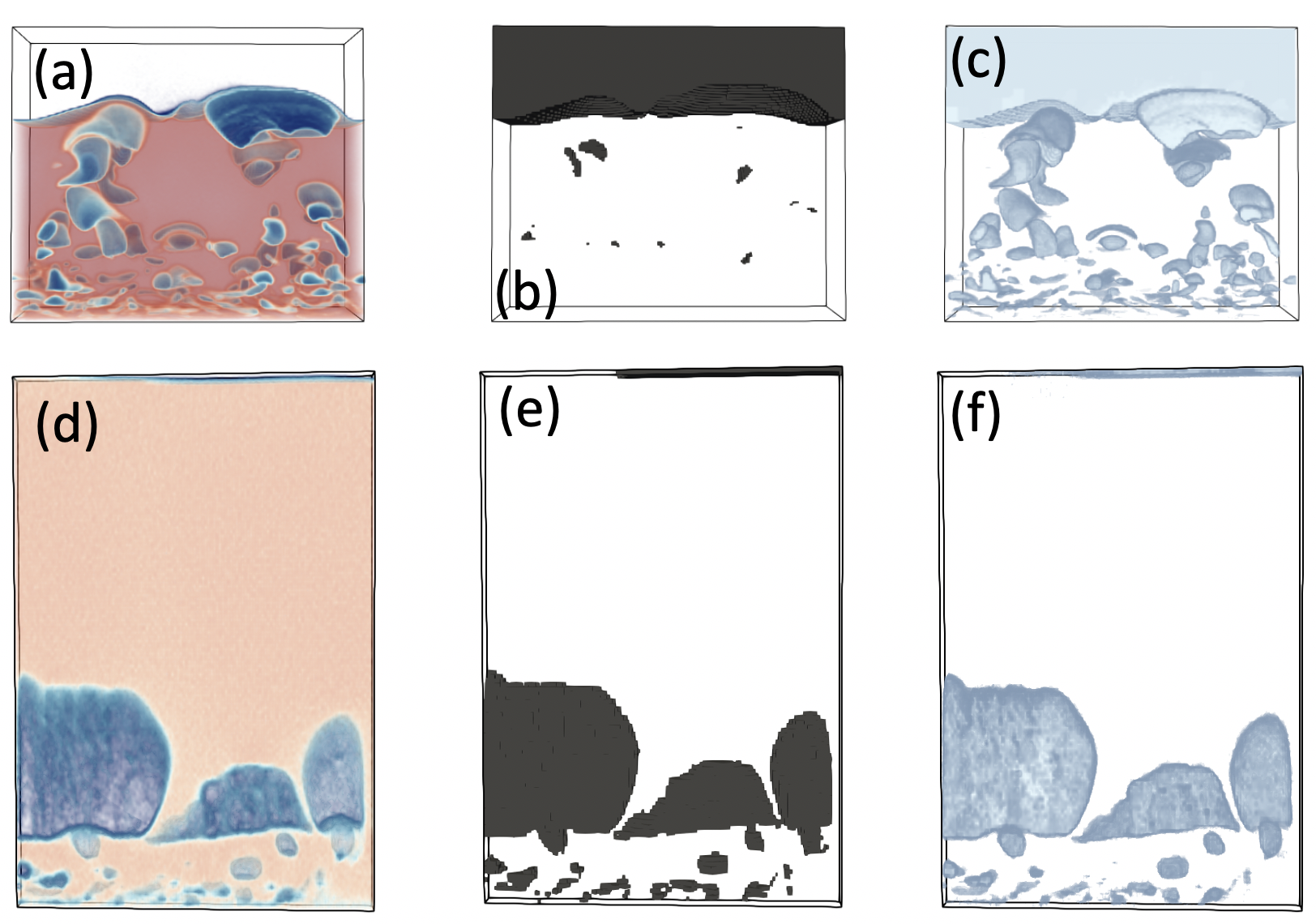}
\caption{Comparing the density threshold-based bubble extraction with the proposed method. The top row ((a), (b), and (c)) and the bottom row ((d), (e), and (f)) show two different simulation use cases. Finding a robust and fixed  density threshold to segment bubbles is not possible since different simulations will have varying density value ranges and the same density threshold will not work. We see that the density threshold=12 woks well for one case, (e), but the same threshold fails to extract bubbles correctly in the second case, (b). In contrast, the proposed statistical method is able to use same feature similarity to extract bubbles as shown in (c) and (f).}
\label{fig:TH_compare}
\end{figure}

\textbf{Parameter Selection and Discussion.}  The cluster size for SLIC is an important parameter. Generally, smaller cluster size improves accuracy since smaller clusters will detect smaller bubble regions more accurately; if the cluster size is too large, some small bubbles will be missed.  Figure~\ref{fig:slic_cluster_param} shows the impact of two different cluster sizes. It can be seen that with large cluster size, some small bubbles are fragmented as highlighted in the red dotted regions. Hence, to capture small bubbles accurately, we have used 3x3x3 cluster sizes in all our experiments. Finally, for segmenting bubbles from the BSFs, we have used a consistent statistical similarity threshold of $0.92$, which resulted in consistent bubble extraction for all the time steps for all simulation use cases. By visually comparing the segmented bubbles with the raw particle fields, it was found that the extracted bubbles accurately represented the void regions. We also found that changing the similarity value slightly does not significantly change the shape of the bubbles and so the classification is robust. Note that, one might just use simple thresholding to extract bubbles directly from the particle density field~\cite{mfix_cise}. However, finding a robust and consistent threshold that will work for different simulation cases is non-trivial since there is no guideline for picking a density threshold, and for different simulation use cases, the dynamic range of density fields will be different. The same threshold will not work across simulations and manual tuning will be required for each simulation which is not desired. In Figure~\ref{fig:TH_compare}, we show that when we use a density threshold = 12 for extracting bubbles, while it works reasonably well in one use case (Figure~\ref{fig:TH_compare}e), the same threshold does not work for the other use case (Figure~\ref{fig:TH_compare}b). In contrast, our statistical method uses the same feature similarity threshold for extracting bubbles for both of the use cases and can extract bubbles accurately as shown in (Figure~\ref{fig:TH_compare}c and Figure~\ref{fig:TH_compare}f). Finally, as the bin frequencies are mapped to particle density values while creating the density field, changing bin numbers will change the values in the density field. We tested several bin numbers: 64x8x64, 128x16x128, 128x24x128, and 256x64x256 for density field generation, finding that 128x16x128 bin numbers produced stable results.

\textbf{Expert Feedback.} Results were presented to the domain experts who are developers of MFIX-Exa and are co-authors of this paper. Overall, the experts were very excited to see the broad and comprehensive capability that we have developed. Before this work, most of the analyses were done offline using visualization tools such as ParaView and were often time-consuming. The experts found the three dimensional bubble visualization capabilities very useful, allowing them to investigate the bubble evolution directly in 3D. The experts appreciated the novel in situ workflow enabling the detection of bubbles in situ followed by the interactive, flexible, and real-time post hoc analysis of bubble dynamics. They agreed that this new analysis capability will accelerate their scientific discovery process significantly while analyzing large-scale simulation data. The experts were comfortable with the Cinema viewers, which consisted of well-known visualization techniques and found them intuitive and useful for interactively exploring bubble dynamics. They felt that the 3D bubble tracking interface worked as a complementary tool to the Cinema viewers, allowing them inspection of bubbles in 3D. Through the PCP, they could filter interesting bubbles based on the various salient characteristics. The addition of SPLOM was also found to be effective for validating relationships between bubble characteristics. Overall, the experts were very positive about our approach and thought that our work has made significant contributions in the analysis of three dimensional bubble dynamics for multiphase flow simulations.

\section{In Situ Case Study}
\begin{figure}[tb] 
\centering
\includegraphics[width=0.9\linewidth]{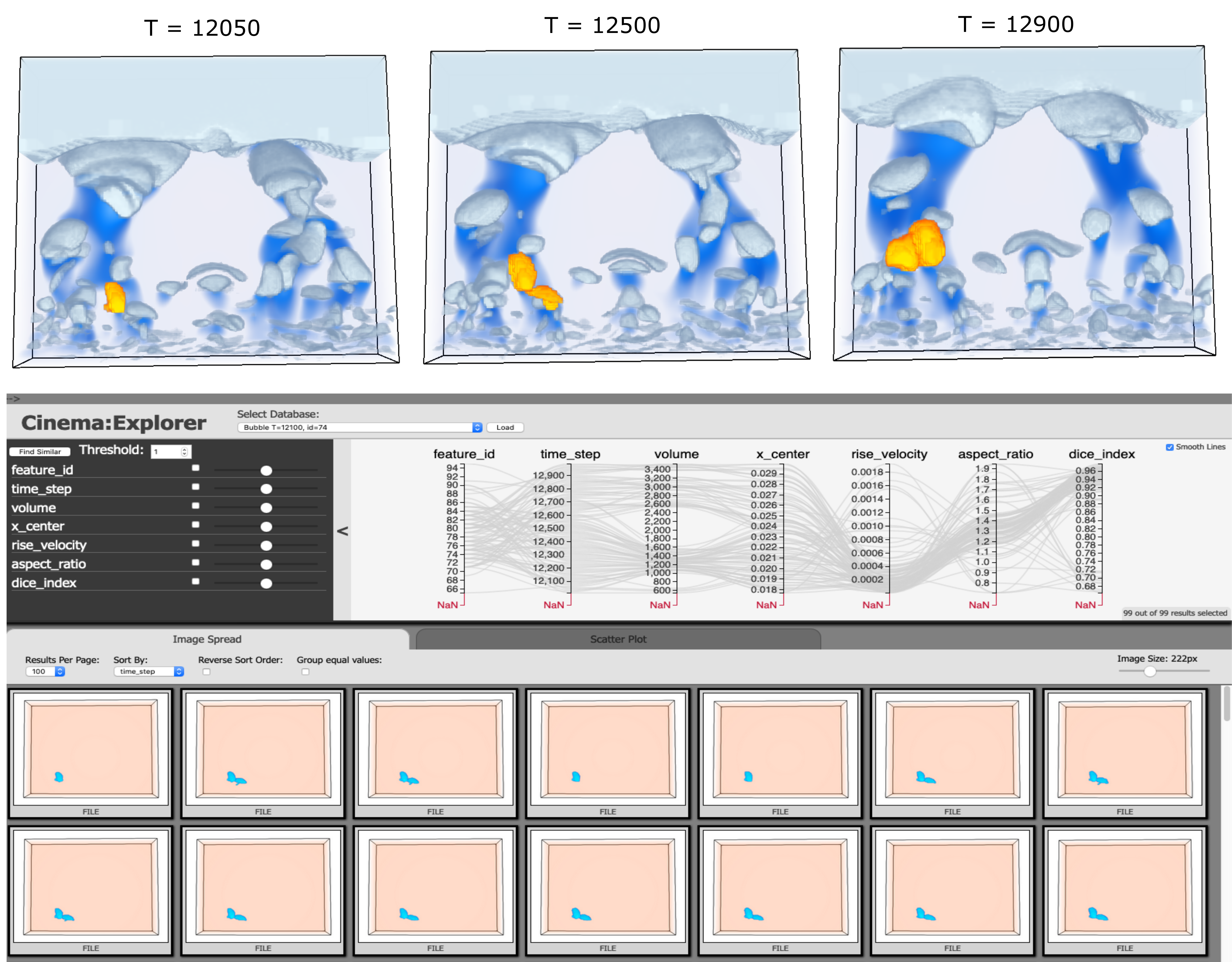}
\caption{Visualization of bubble tracking results from the large-scale in situ MFIX-Exa use case. The feature similarity fields and particle velocity fields shown in the top row are generated in situ.}
\label{fig:new_highres_tracking}
\end{figure}

Until now, we have provided a detailed study of our proposed algorithm and showed its efficacy in exploring bubble dynamics. In this section, we discuss the details of an in situ case study and its computational performance to demonstrate the practicality and in situ viability of our technique. The MFIX-Exa simulation data that we have used above for demonstration purposes, contains $\approx$3 million particles per time step. We have used this data set for developing, benchmarking, and validating our proposed technique by working closely with the MFIX-Exa developers. For the in situ performance study, we have used a significantly larger MFIX-Exa use case. This new use case simulates $54.51$ million particles in a fluidization bed. In this controlled study, we simulate $1000$ time steps for measuring in situ performance and storage benefits of our proposed technique. In Figure~\ref{fig:new_highres_tracking}, we show results from this large-scale simulation where the top row shows 3D visualization of tracking of a specific bubble from three representative time steps and at the bottom, show the CinemaExplorer-based visualization for this bubble. The BSF and the PVFs shown in the top row are generated in situ and our tracking interface is used to generate these visualizations. Since our method only requires particle location and velocity, we have stored only the particle positions and velocity while writing the data out for a fair storage reduction comparison.

\subsection{In Situ Code Integration}
The MFIX-Exa~\cite{mfix_exa} and AMReX~\cite{amrex_paper} codes are developed in C++. To perform in situ data analysis, custom code was added to both MFIX-Exa and AMReX code bases. The in situ code is  developed in C/C++ and uses VTK data models~\cite{VTK}. The final summarized bubble similarity fields, and the particle rise velocity fields are stored in VTK image data format, and so, can be readily loaded into standard visualization tools such as ParaView~\cite{Paraview} or VisIt~\cite{visit}. A direct data access scheme is used to read the raw particle data from memory by querying the particle container data structures of AMReX.  Since the simulation and the in situ data processing routines use the same memory and computing resources, the proposed in situ integration works in synchronous mode, tightly coupled with the simulation.  As we output the in situ summarized feature-specific data in the form of 3D scalar fields that can be analyzed and visualized to produce new results, the output type of our in situ processing is explorable.

\subsection{In Situ Performance Study}
The in situ performance study was done on the Summit supercomputer~\cite{summit}, an IBM system located at the Oak Ridge Leadership Computing Facility. Each compute node of Summit contains two IBM POWER9 processors, $512$ GB of DDR4 memory, $1.6$~TB of non-volatile memory, and six NVIDIA Tesla V$100$ accelerators.

\textbf{Storage savings.}
A single time step of this large-scale simulation outputs $3.1$~GB of raw data containing particle id, location, and velocity. Note that this simulation typically stores many other fluid variables, which will increase the storage further. However, since we are only using particle location and velocity variables, we only store those variables so that we get a fair storage comparison. We simulated $1000$ time steps, producing over 3~TBs of data. Due to the high data volume, experts often only store every 50th or every 100th time step so the resultant data size remains tractable. In contrast, processing data in situ allows us to access higher temporal fidelity so that more time-accurate information about the bubbles can be captured. In our experiment, we accessed simulation data every $10$th time step for $1000$ total time steps, resulting in storage of $100$ time steps. Storing raw data for these $100$ time steps requires $310$~GB storage space. Instead, by performing in situ analysis, for every $10$th time step, we computed bubble similarity and particle rise velocity fields and stored those fields to disk. The disk space required for storing the in situ generated data is only $224$~MB in VTK format which is significantly smaller compared to the raw particle data storage. Besides the raw particle data, the simulation also needs to store checkpoint restart files that are significantly larger. For our use case, if we store checkpoint files at every $50$th time step, it results in an additional $271$~GB of storage. Further, note that, these simulations are expected to run for significantly longer duration when initialized with unknown input conditions, producing tens of thousands of time steps and admittedly, storing raw particle data for such long runs is going to be prohibitive. Hence, an in situ analysis workflow will be critical for scientific insight and will result in a significant data triage while paving a path for flexible and detailed post hoc study of bubble dynamics.

\begin{table}[t]
\centering
\caption{In situ processing times (in seconds) taken by the proposed method compared to the simulation time.} \label{insitu_table_1}
\resizebox{\columnwidth}{!}{%
\begin{tabular}{|c|c|c|c|c|c|}
\hline
 & \begin{tabular}[c]{@{}c@{}}Density and  \\ velocity field\end{tabular} & SLIC & \begin{tabular}[c]{@{}c@{}}Similarity\\ field\end{tabular} & \begin{tabular}[c]{@{}c@{}}Total in situ\\ computation\end{tabular} & Simulation \\ \hline
512 processes & 4.39 & 122.41 & 1.45 & 128.25 & 8735.09 \\ \hline
1024 processes & 3.45 & 121.3 & 1.44 & 126.19 & 5528.79 \\ \hline
2048 processes & 2.58 & 124.37 & 1.44 & 128.39 & 4408.6 \\ \hline
4096 processes & 2.44 & 123.20 & 1.46 & 127.1 & 4114.35 \\ \hline
\end{tabular}
}
\end{table}

\begin{table}[tb]
\centering 
\caption{Post hoc timings (in seconds) for different steps of our proposed algorithm. By processing data in situ, timings shown in this table can be saved.} \label{offline_table}
\scalebox{0.9}{%
\begin{tabular}{|c|c|c|c|}
\hline
\begin{tabular}[c]{@{}c@{}}Density and\\ velocity field\end{tabular} & SLIC & \begin{tabular}[c]{@{}c@{}}Similarity\\ field\end{tabular} & Total I/O \\ \hline
39420.87 & 59.27 & 144.57 & 3364.43 \\ \hline
\end{tabular}
}
\end{table}

\begin{figure}[!tb]
\centering
\includegraphics[width=0.8\linewidth]{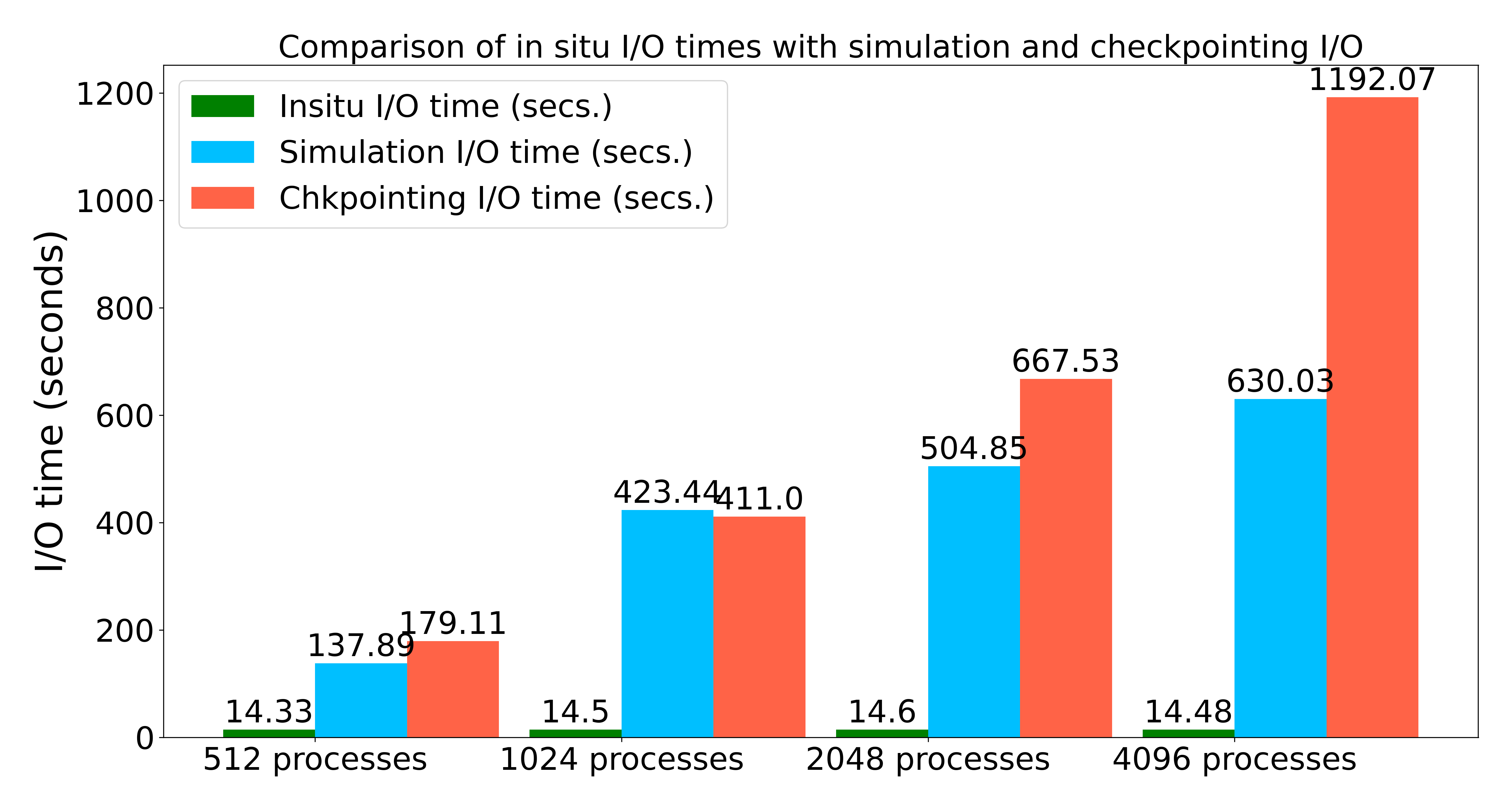}
\caption{Comparison of in situ I/O times with simulation and check-pointing I/O times. We see that the in situ I/O is significantly lower compared to the simulation and check-pointing I/O.}
\label{fig:IO_plot}
\end{figure}

\textbf{Computational time savings.}
Since the simulation starts from a random particle initialization and gradually reaches a state where bubbles are periodically formed, we started MFIX-Exa simulation from a later time point (time step = $12000$) to produce bubbles and simulated $1000$ time steps. In Table~\ref{insitu_table_1}, we provide the computational timings taken by the different steps of the algorithm. The in situ computation of particle density field and PVF require distributed communication and we see that as we increase the number of MPI processes, this part of the algorithm scales (first column of Table~\ref{insitu_table_1}). Also, the computation of SLIC and similarity field requires no data communication and is done by a single process. So we observe that they take a constant fraction of time. When we observe the total in situ processing time and compare it to the simulation time, we find that the in situ processing is taking only a small percentage of the simulation time as desired and is not bottle-necking the simulation. In Table~\ref{offline_table}, we show the time it would take if the same computation were done offline using raw particle data on a standard desktop computer. We see that the off-line processing of such a large data set takes a significantly longer time compared to the time taken in situ (Table~\ref{insitu_table_1}). The density and velocity field computation time takes the largest fraction of computation time since this computation accesses the raw particle data. The computation of SLIC and similarity field uses the derived particle density field which is much smaller than the raw data and hence the computation is faster. We can use parallel processing to improve upon these computation timings, however, the I/O time (third column of Table~\ref{offline_table}) will still be high for post hoc analyses, and as the number of time steps to process will increase, these post hoc computation timings will become significantly high. In contrast, by processing data in situ, we can completely bypass timings reported in Table~\ref{offline_table} and the domain experts can start their bubble analysis right after the simulation ends and the analysis will be accelerated. 

Figure~\ref{fig:IO_plot} shows the advantage of in situ processing for I/O reduction. If we store the raw particle data (at every 10th time step) and checkpoint files (at every 50th time step), we see that the I/O time taken is significantly higher and increases with the number of MPI processes. This I/O bottleneck is the primary concern of the scientists. By processing data in situ, we reduce the size of the output data, and as a result, the in situ I/O is significantly smaller (the green bars) compared to the simulation and checkpoint I/O (blue and orange bars). Hence, we demonstrate that in situ processing results I/O and computation time reduction and provides a practical pathway for analyzing very large-scale simulation data sets.

\section{Conclusions}
In this work, we have presented an end-to-end in situ analysis guided visual exploration of complex bubble dynamics phenomena in fluidized bed simulations. We have successfully demonstrated an in situ analysis pipeline with MFIX-Exa to extract bubble specific data in situ, enabling flexible and scalable post hoc exploration of bubble dynamics. We have conducted a detailed performance study, and validated our findings through domain expert feedback and through qualitative comparison with existing literature in the multiphase flow analysis. In the future, we plan to apply our technique to ensemble MFIX-Exa simulations so that the role of input parameters in the bubble dynamics can be explored efficiently. We also want to explore other particle density estimation techniques provided in~\cite{tom_density} for computing particle density fields. Finally, GPU implementations of our in situ algorithm are underway to further improve in situ  performance of future extreme-scale MFIX-Exa runs with hundreds of millions of particles at upcoming exascale supercomputers~\cite{mfix_ecp_url}.



\ifCLASSOPTIONcompsoc
  \section*{Acknowledgments}
\else
  \section*{Acknowledgment}
\fi

The authors would like to thank the Department of Energy and Los Alamos National Laboratory for the funding and support in carrying out this research. This research was supported by the Exascale Computing Project (17-SC-20-SC), a collaborative effort of the U.S. Department of Energy Office of Science and the National Nuclear Security Administration and is released under LA-UR-21-23312. This research used resources of the Oak Ridge Leadership Computing Facility at the Oak Ridge National Laboratory, which is supported by the Office of Science of the U.S. Department of Energy under Contract No. DE-AC05-00OR22725.

\ifCLASSOPTIONcaptionsoff
  \newpage
\fi



%

\bibliographystyle{IEEEtran}
\bibliography{insitu-bubbles}

%

\begin{IEEEbiography}[{\includegraphics[width=1in,height=1.25in,clip,keepaspectratio]{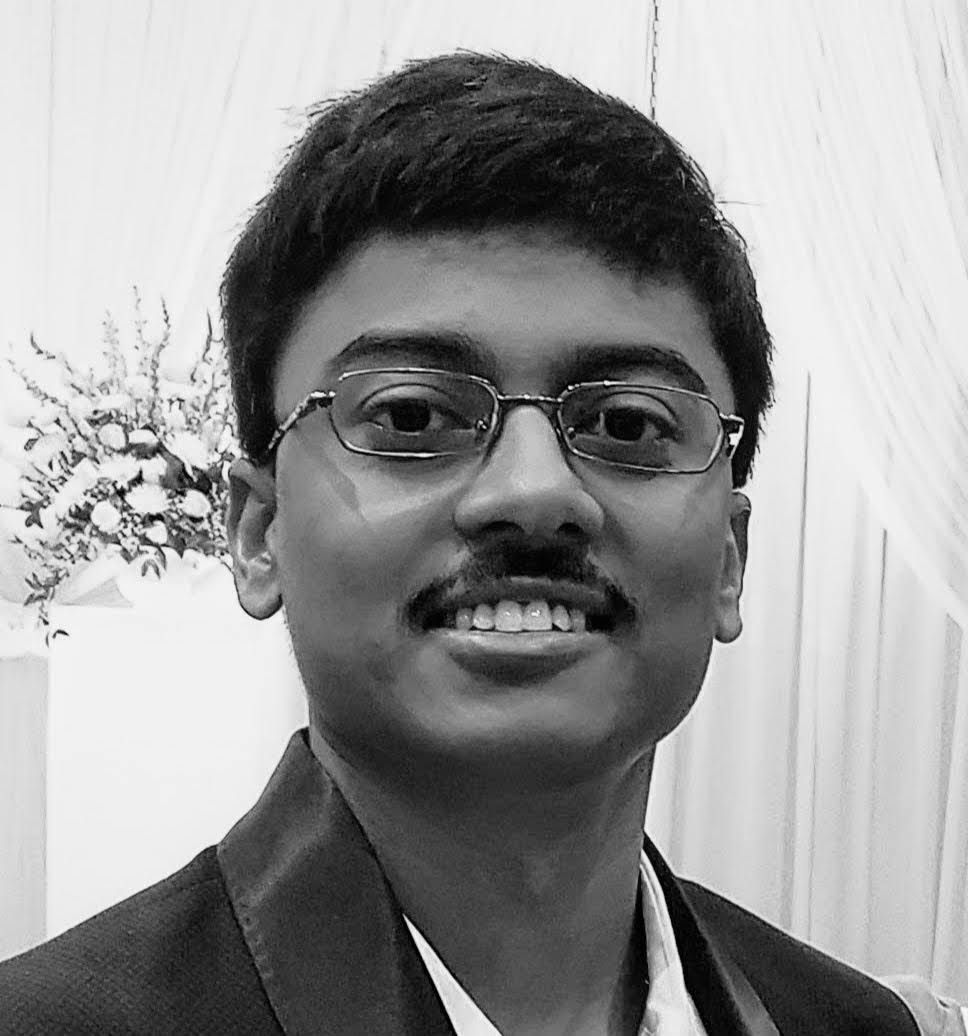}}]{Soumya Dutta}
is a staff scientist in Data Science at Scale team at Los Alamos National Laboratory. He received his M.S. and the Ph.D. degree in Computer Science and Engineering from the Ohio State University in May 2017 and May 2018 respectively. His research interests are statistical data summarization and analysis; in situ data analysis, reduction, and feature exploration; machine learning and uncertainty analysis; and time-varying, multivariate data exploration. He is a member of the IEEE and the IEEE Computer Society. Contact him at sdutta@lanl.gov.
\end{IEEEbiography}

\vskip 0pt plus -1fil

\begin{IEEEbiography}[{\includegraphics[width=1in,height=1.25in,clip,keepaspectratio]{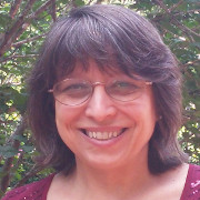}}]{Terece L. Turton}
is a staff scientist in Information Sciences at Los Alamos National Laboratory. Her current research interests include in situ workflows for exascale simulations, scientific visualization, and perceptual user evaluation.  In addition to her research roles, she provides technical project management for the Exascale Computing Project's ALPINE/ZFP project. She received her Ph.D. in Physics from the University of Michigan in 1993. She is a member of the IEEE and the IEEE Computer Society. Contact her at tlturton@lanl.gov.
\end{IEEEbiography}

\vskip 0pt plus -1fil

\begin{IEEEbiography}[{\includegraphics[width=1in,height=1.25in,clip,keepaspectratio]{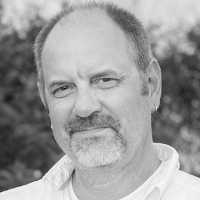}}]{David H. Rogers}
is the team lead of the Data Science at Scale Team at Los Alamos National Lab. He joined LANL in 2013, after a decade of leading the Scalable Analysis and Visualization Team at Sandia National Labs, where he was instrumental in bringing in-situ analysis and visualization into production. He now focuses on interactive analysis tools that integrate design, scalable analytics and principles of cognitive science to promote scientific discovery. Prior to working on large scale data analysis, David worked at DreamWorks Feature animation, writing and managing production software. 
He has degrees in Computer Science, Architecture (buildings, not computers), and an MFA in Writing for Children. Contact him at dhr@lanl.gov.
\end{IEEEbiography}

\vskip 0pt plus -1fil

\begin{IEEEbiography}[{\includegraphics[width=1in,height=1.25in,clip,keepaspectratio]{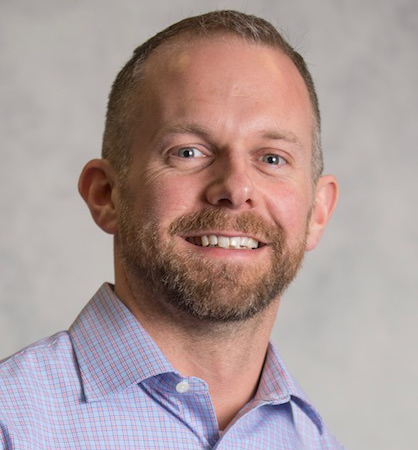}}]{Jordan M. Musser}
is a physical research scientist at the National Energy Technology Laboratory. He received his Ph.D. from West Virginia University. His research interests include the development and application of computational multiphase models, computational methods for dense particle laden flows, and their application to large industrial systems. Contact him at jordan.musser@netl.doe.gov.
\end{IEEEbiography}

\vskip 0pt plus -1fil

\begin{IEEEbiography}[{\includegraphics[width=1in,height=1.25in,clip,keepaspectratio]{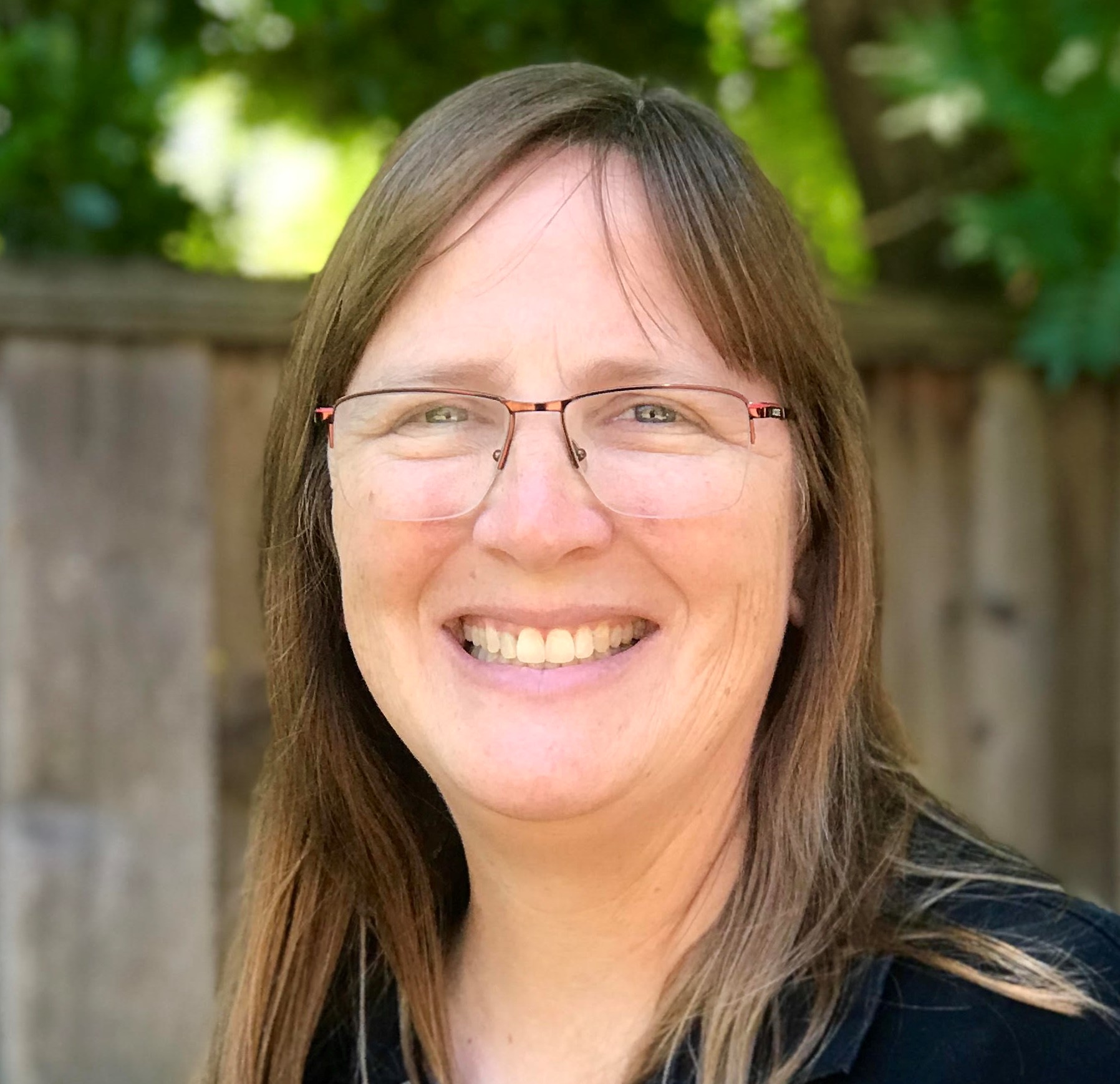}}]{Ann S. Almgren}
is a senior scientist in the Computational Research Division of Lawrence Berkeley National Laboratory and the Group Lead of the Center for Computational Sciences and Engineering. Her primary research interest is in computational algorithms for solving PDEs in a variety of application areas. Her current projects include the development and implementation of new multiphysics algorithms in high-resolution adaptive mesh codes designed for the latest hybrid architectures. She is an SIAM Fellow and the Deputy Director of the ECP AMR Co-Design Center. Contact her at asalmgren@lbl.gov.
\end{IEEEbiography}

\vskip 0pt plus -1fil

\begin{IEEEbiography}[{\includegraphics[width=1in,height=1.25in,clip,keepaspectratio]{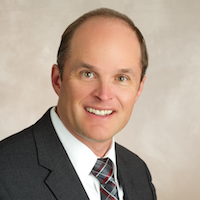}}]{James P. Ahrens} of Los Alamos National Laboratory (LANL) is the founder and design lead of ParaView, a widely adopted visualization and data analysis package for large-scale scientific simulation data. Dr. Ahrens graduated in 1996 with a Ph.D. in computer science from the University of Washington. Following his graduate studies, he joined LANL as a technical staff member. Dr. Ahrens has published over 100 peer-reviewed papers that have been cited over 5000 times. Dr. Ahrens is the lead of the LANL Information Science and Technology Institute, and the lead of the Data and Visualization area for the Exascale Computing Project.
\end{IEEEbiography}




\end{document}